\documentclass[aps,prb,onecolumn,longbibliography]{revtex4-2}

\usepackage{amsmath,amssymb,mathrsfs}
\usepackage{amsbsy,amsthm,amsfonts}
\usepackage{color}
\usepackage[dvipsnames]{xcolor}
\usepackage[colorlinks,citecolor=blue]{hyperref}
\usepackage{bm}
\usepackage{graphics,graphicx}

\newcommand\figref[1]{Fig.~\ref{#1}}

\newcommand\sectref[1]{Sec.~\ref{#1}}

\newcommand{\bfe}   {\mathbf{e}}

\newcommand{\bfh}   {\mathbf{h}}

\newcommand{\K}  {k}  

\newcommand{\calD}    {\mathcal{D}}

\newcommand{\DtN}  {\mathcal{D}}

\begin{document}
	
	\title[2D Bulk-Boundary Correspondence]{Bulk-Boundary Correspondence in 2D Topological Photonics: Analysis and Simulation}
	\author{Igor Tsukerman}
	
	\affiliation{Department of Electrical and Computer Engineering, 
		The	University of Akron, OH 44325-3904\\
		E-mail: igor@uakron.edu}
	
	% use {asbstract*} to suppress the copyright line. Copyright information will be added in production
	
	\begin{abstract} 
The centerpiece  of topological photonics is the bulk-boundary correspondence principle (BBCP), which relates discrete invariants of the Bloch bands to the possible presence of interface modes between two periodic heterostructures. In addition to the fundamental significance of BBCP, interface modes per se are of interest in a variety of applications. In Maxwell's electrodynamics, the BBCP has been rigorously proved for 1D problems, but
%and ``1.5D'' problems (in the latter, fields and wave vectors generally have two components, while the mathematical equations can still be reduced to 1D). 
the 2D case is qualitatively different, as the boundary conditions involve nontrivial Dirichlet-to-Neumann maps rather than scalar impedances as in 1D. The theoretical analysis and numerical examples in the paper are consistent with the BBCP. Moreover, the BBCP is closely connected with the positivity of electromagnetic energy density, as has also been shown to be true in 1D cases.
	\end{abstract}

	\date{\today}

	\maketitle
	
%%%%%%%%%%%%%%%%%%%%%%%%%%  body  %%%%%%%%%%%%%%%%%%%%%%%%%%

%%%%%%%%%%%%%%%%%%%%%%%%%%%%%%%%%%%%%%%%%%%%%%%%%%%%%%%%%%
\section{Introduction}\label{sec:Introduction}
%%%%%%%%%%%%%%%%%%%%%%%%%%%%%%%%%%%%%%%%%%%%%%%%%%%%%%%%%%
%
The main tool of field analysis in periodic structures is propagating Bloch modes, forming a set of bands separated by frequency gaps. Under assumptions (i)--(ii)  below, each band can be assigned a discrete \textit{topological invariant} -- Zak phase, zero or $\pi$, in 1D or an integer Chern number in 2D \cite{Haldane-Raghu-PRL08,Wang-Chong-Joannopoulos-PRL2008,Vanderbilt-Berry-phases-book2018}. The topological invariance is with respect to any continuous bandgap-preserving changes of geometric and material parameters of the lattice cell. Throughout the paper, we assume that (i) the lattice cells are mirror-symmetric (\sectref{sec:Formulation}), and (ii) losses can be neglected, so that non-Hermitian topological photonics \cite{Banerjee-Non-Hermitian2023,Yan-Gong-NonHermitian-topo-review} remains outside the scope of the paper.

If two different periodic structures abut, there may or may not exist modes confined to the interface. The centerpiece of topological photonics is the bulk boundary correspondence principle (BBCP), which relates the possible presence %, and number, of topologically protected 
of interface modes in a bandgap to the discrete invariants of Bloch bands. In addition to its fundamental significance, BBCP has practical applications, since topologically protected modes are of interest in communication, topological lasers, and other areas \cite{Wang-Chong-Joannopoulos-PRL2008,Wang-Chong-Joannopoulos-Nature2009,
	%Khanikaev-Shvets-photonic-topological13,	
	Khanikaev-Shvets-2D-topological17,%Ma-Shvets-topological-waveguides15,
	Xue-Yang-Zhang-Valley-photonics21}. 

The BBCP is surprising in several respects. First, note that Bloch bands can be defined and calculated over a single lattice cell under Bloch-periodic boundary conditions (\sectref{sec:Formulation}); this has nothing to do with material interfaces, where the translational symmetry of an infinite lattice is broken. Thus, conclusions about \textit{boundary} modes are drawn after ignoring the presence of the boundary to begin with; this is puzzling.

Furthermore, the BBCP relates \textit{evanescent} modes within a particular bandgap\footnote{More precisely, within a frequency overlap of the bandgaps of the two adjacent heterostructures.} to  \textit{propagating} waves \textit{at all lower frequencies}. Indeed, the relevant parameter is the \textit{gap} Chern number or \textit{gap} Zak phase -- that is, the sum of Chern numbers (Zak phases) over all bands below a given gap. Somehow, the existence of \textit{evanescent} boundary modes at, e.g., optical frequencies depends on \textit{propagating} waves at radio frequencies, power frequencies, and even on the static field. This is perplexing and needs to be understood from both mathematical and physical perspectives.

A short historical note on the status of the BBCP is apposite. The relevant developments span over half a century -- from Pancharatnam's work in the 1950s \cite{Pancharatnam56} to the quantum Hall effect \cite{vonKlitzing-QH-effect1980} to the discovery of topological insulators \cite{Hasan-Kane-Colloquium-TI2010}, with many advances in between. The findings and conclusions of the elaborate topological theories in condensed matter physics were then translated to electrodynamics \cite{Haldane-Raghu-PRL08,Wang-Chong-Joannopoulos-PRL2008,Wang-Chong-Joannopoulos-Nature2009}. While this translation has been quite productive and has resulted in vigorous research activity over the last two decades, there are significant differences between the equations and boundary conditions of condensed matter physics and electrodynamics. Many topological theories of the former involve tight binding, lattice or finite difference models \cite{Hatsugai-PRL1993,Hatsugai-PRB1993,%Thiang-Topological-states-2020,
	Prodan-Schulz-Baldes-2016}.
%Bianco-Resta-Topological-order2011,Caio-Topological-marker-currents-2019}, 
% while Maxwell's electrodynamics has to do with  waves in continuous media. 
In the continuous case, there exists sophisticated mathematical work utilizing, among other things, Fredholm index theory for  single-particle Schr\"{o}dinger Hamiltonians % in the context of the quantum Hall effect
\cite{Elbau-Graf-Bulk-edge-conductance-2002,Kellendonk-Schulz-Baldes-edge-currents-2004,Drouot-Microlocal-2021}; it remains to be seen to what extent these studies apply to electromagnetic problems.

The present paper deals exclusively with continuous Maxwell electrodynamics. 
%In 2008, Wang \textit{et al.} remarked \cite{Wang-Chong-Joannopoulos-PRL2008}: ``The Chern-number argument for the existence of photonic edge modes depends on a crucial assumption: that Hatsugai's relation between edge states and Chern numbers ..., derived using a lattice ... model, applies to the photonic system. ... this has not been formally proven''. 
Over the last ten years, the BBCP was rigorously proved for lossless media in 1D and 1.5D \cite{Xiao-Chan-Geom-phases-1D-2014,Thiang-Zhang-Bulk-interface-1D-2023,Tsukerman-Markel-EPL2023,Coutant-Lombard-impedance-topology-2024}. ``1.5D'' refers to the case where material parameters depend only on the coordinate $n$ normal to the interface; mathematically, the problem can be reduced to one dimension, but the fields and wave vectors may have two components.  
A central role in the analysis and proof of the BBCP in 1D is played by the Bloch boundary impedance or, alternatively, admittance (the inverses of one another) \cite{Xiao-Chan-Geom-phases-1D-2014,Thiang-Zhang-Bulk-interface-1D-2023,Tsukerman-Markel-EPL2023,Coutant-Lombard-impedance-topology-2024}. In 2D, the scalar admittance turns, mathematically, into an  infinite dimensional
(in the functional-analytic sense) Dirichlet-to-Neumann (DtN) map (\sectref{sec:DtN}).
%, which can be expressed, e.g., in the basis of spatial harmonics along the interface.

The paper is devoted to an analytical, semi-analytical and numerical study of the DtN maps and the respective interface matching conditions. Four interrelated Bloch and Dirichlet problems are formulated in \sectref{sec:Formulation}. The DtN map for a semi-infinite strip can be defined and computed via a set of evanescent Bloch modes in the lattice cell (\sectref{sec:DtN}). Numerical case studies include a combination of gyromagnetic and simple dielectric/magnetic media (\sectref{sec:Examples}).
%
%%%%%%%%%%%%%%%%%%%%%%%%%%%%%%%%%%%%%%%%%%%%%%%%%%%%%%%%%%%%%%%%%%%%%%%%%%%%%%%
\section{Formulation of four related problems}\label{sec:Formulation}
\begin{figure}
	\centering
	\includegraphics[width=0.5\linewidth]{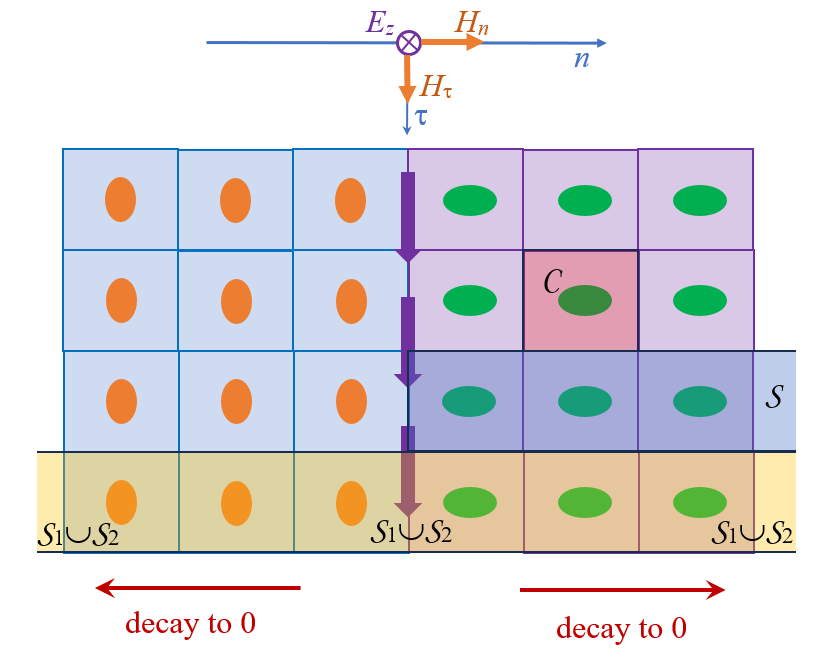}
	\caption{An illustration of two abutting structures; $s$ mode is assumed.
	The purple vertical arrows indicate a possible direction of propagation of an interface mode.
%	, if it exists.
}
	\label{fig:setup-2D-smode}
\end{figure}

%%%%%%%%%%%%%%%%%%%%%%%%%%%%%%%%%%%%%%%%%%%%%%%%%%%%%%%%%%%%%%%%%%%%%%%%%%%%%%%
%
A canonical setup is shown in \figref{fig:setup-2D-smode}. In this paper, only rectangular lattices are considered, with cells scaled to $a \times a$ squares for convenience. Each of the two periodic heterostructures fills its respective half-space. A fundamental question is this: Under what conditions do there exist one or more modes evanescent in the normal direction $n$ -- either standing waves or traveling in the tangential direction $\tau$ along the boundary? Purportedly, the BBCP answers this question; however, as noted in the Introduction, a rigorous proof exists in some cases but not, to the best of my knowledge and understanding, for 2D Maxwell problems.

The main part of the paper focuses on the purest case of topological electrodynamics, whereby losses and frequency dependence (``frequency dispersion'') of material parameters are neglected. The lossy case falls under the umbrella of non-Hermitian photonics \cite{Yan-Gong-NonHermitian-topo-review}
		and, in 1D, is the subject of Ref.~\cite{Felbacq-Rousseau24}. Accounting for frequency dispersion in my analysis is not conceptually difficult (see Appendix); the relevant techniques have already been worked out in 1D \cite{Coutant-Lombard-impedance-topology-2024} and in 1.5D  \cite{Tsukerman-Markel-Topology-Bloch-impedance2024}.

For definiteness, we assume the $s$-mode (one-component electric field $\bfe = \hat{z} e$ and the respective two-component magnetic field $\bfh = h_n \hat{n} + h_{\tau} \hat{\tau}$); analysis for the $p$-mode is similar. Hats indicate  unit vectors; the coordinate system $(n, \tau, z)$ is right-handed.
The complex amplitudes $\bfe(n)$ and $\bfh(n)$ satisfy Maxwell's equations:
\begin{equation}\label{eqn:Maxwell}
	\nabla \times \bfe(n, \tau) = i  \K \mu(n, \tau) \, \bfh(n, \tau),
	\quad
	\nabla \times \bfh(n, \tau) = -i  \K \epsilon(n, \tau) \, \bfe(n, \tau)
\end{equation}
in the Gaussian system under the $\exp(-i \omega t)$ phasor convention; $\K = \omega / c$ is the wave number. For the $s$ mode, the magnetic permeability $\mu(n, \tau)$ may in some cases be anisotropic and represented in the $(n, \tau)$ coordinates as a $2 \times 2$ matrix (\sectref{sec:Examples}); $\epsilon$ is scalar.
Eliminating the magnetic field $\bfh$ from \eqref{eqn:Maxwell}, one arrives at the electric field equation:
\begin{equation}\label{eqn:div-rho-grad-e-k2-eps-e}
	\nabla \cdot \left( \rho(n, \tau) \nabla  e(n, \tau) \right) + \K^2 \epsilon(n, \tau) e(n, \tau) ~=~ 0
\end{equation}
where $\rho$ is related to the magnetic permeability tensor $\mu$ and its inverse $\nu$ as
\begin{equation}\label{eqn:rho-nu-mu}
	\rho = \begin{pmatrix}
		     \rho_{n n}  &  \rho_{n \tau}\\
		     \rho_{\tau n}  &  \rho_{\tau \tau}
  	        \end{pmatrix} ~=~ 
  	        \begin{pmatrix}
  	        \nu_{\tau \tau} &  -\nu_{\tau n}\\
  	        -\nu_{n \tau} & \nu_{n n}
  	        \end{pmatrix};
  	        \quad \quad
  	        \nu = 
  	        \begin{pmatrix}
  	          \nu_{n n}  &  \nu_{n \tau}\\
  	           \nu_{\tau n}  &  \nu_{\tau \tau}
  	        \end{pmatrix}
  	        ~=~ \mu^{-1}
\end{equation}
For a given periodic heterostructure on a rectangular lattice, the material parameters satisfy
\begin{equation}\label{eqn:periodicity}
	\epsilon(n + m_n a, \tau + m_{\tau} a) ~=~ \epsilon(n, \, \tau);
	\quad
	\mu(n + m_n a, \tau + m_{\tau} a) ~=~ \mu(n, \, \tau);	
	~~ (m_n, m_{\tau}) \in \mathbb{Z}^2
\end{equation}
as long as both points $(n + m_n a, \tau + m_{\tau} a)$ and $(n, \, \tau)$ lie within that structure.
For brevity, and without much loss of generality, we assume that the medium right next to the interface boundary is nonmagnetic: $\rho(n=0, \tau)$ = 1.
The lattice cells are assumed to be mirror-symmetric with respect to $n$:
\begin{equation}\label{eqn:mirror-symmetry}
	\epsilon(a-n, \tau) ~=~ \epsilon(n, \, \tau);
	\quad
	\mu(a-n, \tau) ~=~ \mu(n, \, \tau)
\end{equation}
Note that if the structure occupies the whole space, mirror symmetry and periodicity imply that the material parameters are even functions of $n$:
\begin{equation}\label{eqn:even-functions}
	\epsilon(-n, \tau) ~\overset{\text{periodicity}}{=}~ \epsilon(a-n, \, \tau) 
	~\overset{\text{symmetry}}{=}~ \epsilon(n, \, \tau)
\end{equation}
and the same for $\mu$.
The overall setup  is schematically represented in \figref{fig:setup-2D-smode}. The boundary modes satisfy Maxwell's equations \eqref{eqn:Maxwell}, \eqref{eqn:div-rho-grad-e-k2-eps-e}, with Bloch boundary conditions \eqref{eqn:Bloch-conditions-tau} on the $\tau$ sides of a single semi-infinite strip $S = [0, \infty] \times [-a/2, a/2]$, and decaying to zero in the normal direction away from the interface in both structures.

Four interrelated problems need to be considered. 
One is the Bloch problem over a single lattice cell $C: 0 \leq n \leq a; ~ -a/2 \leq \tau \leq a/2$:
\begin{equation}\label{eqn:Bloch-conditions-n}
	e(n+a, \, \tau) = \exp(i q a) \, e(n, \, \tau);
	\quad \quad
	\partial_n e(n+a, \, \tau) = \exp(iqa) \, \partial_ne(n, \, \tau) 
\end{equation}
\begin{equation}\label{eqn:Bloch-conditions-tau}
	e(n, \, \tau+a) = \exp(i \beta a) \, e(n, \, \tau);
	\quad \quad
	\partial_\tau e(n, \, \tau+a) = \exp(i \beta a) \, \partial_\tau e(n, \, \tau)
\end{equation}
Different symbols for the Bloch wavenumbers $q$ (in the $n$ direction) and $\beta$ (in the $\tau$ direction) are used because they play different roles in the physical setup and mathematical analysis. In particular, of our primary interest are modes propagating in the $\tau$ direction with a real $\beta$ and evanescent in the $n$ direction, with a purely imaginary $q$. The Maxwell cell problem \eqref{eqn:Maxwell} with Bloch boundary conditions \eqref{eqn:Bloch-conditions-n}, \eqref{eqn:Bloch-conditions-tau} has three eigenparameters: $\K$, $\beta$ and $q$. It will be convenient to treat the first two of them as given and the third one, $q$, as an unknown. Some expressions simplify if the $\tau$-periodic factor $\tilde{e}(n, \, \tau)$ is used instead of $e(n, \, \tau)$:
\begin{equation}\label{eqn:e-eq-tilde-e-exp}
	e(n, \, \tau) = \tilde{e}(n, \, \tau) \exp(i \beta \tau);
	\quad
	\tilde{e}(n, \, a/2) = \tilde{e}(n, \, -a/2)
\end{equation}
The second and third problems (\sectref{sec:DtN}) have to do with the DtN operators on the boundary of a single lattice cell and a semi-infinite strip $S$, respectively; the strip is composed of an infinite number of cells. Formally, the boundary conditions for $S$ are \eqref{eqn:Bloch-conditions-tau} on the $\tau = \pm a/2$ sides and
\begin{equation}\label{eqn:e-0-at-infty}
	e(n, \, \tau) \rightarrow 0 \text{~as~} n \rightarrow \infty;
	\quad \quad
	\partial_n e(n, \, \tau) \rightarrow 0 \text{~as~} n \rightarrow \infty
\end{equation}
uniformly with respect to $n$, $\tau$.
One complication is that the standard radiation conditions at $n \rightarrow \infty$ do not apply, since there is no free space in this setup. Instead, the condition at infinity can be defined via a set of evanescent Bloch modes in a lattice cell:
\begin{equation}\label{eqn:bc-infinity}
	e(n, \tau) = \sum_m c_m e_m(n, \tau);
	\quad
	\sum_m |c_m|^2 = 1
\end{equation}
%
%where
%
\begin{equation}\label{eqn:evanescent-Bloch-modes}
	e_m(n, \tau) ~
	\text{satisfy}~\eqref{eqn:Maxwell}, ~ \eqref{eqn:Bloch-conditions-n},
 ~ \eqref{eqn:Bloch-conditions-tau},
	~\text{with}~ q = q_m, ~~ m = 1,2, \ldots;
	~~
	\int_C (e_m^2 + |\nabla e_m|^2) \, dC = 1
\end{equation}
Strictly speaking, \eqref{eqn:evanescent-Bloch-modes} for each $m$ defines a family of modes which differ by an arbitrary phase factor $\exp(i \phi)$, $\phi \in \mathbb{R}$; let $e_m$ denote one representative of this family.

Finally, solutions over the abutting semi-infinite strips $S_{1,2}$ must be matched at the interface through Maxwell’s boundary conditions (\sectref{sec:DtN}). 
%
%\begin{equation}\label{eqn:symmetry-operator}
%	\mathcal{P}_n e(n, \tau) \overset{\text{def}}{=} e(-n, \tau)
%\end{equation}
%
%
%The parity of Bloch modes at the $\Gamma$ and $X$ points is known to have major implications in 1D, through a direct connection with the nulls and poles of the Bloch boundary impedance (see \cite{Tsukerman-Markel-EPL2023} for details and references). 

It is proved in \cite{Thiang-Zhang-Bulk-interface-1D-2023,Tsukerman-Markel-EPL2023,Coutant-Lombard-impedance-topology-2024} that Bloch impedance monotonically decreases as a function of frequency within any band gap. 
In 2D, the situation is qualitatively more involved. Admittance turns into a DtN map, defined below and well known in the theory of partial differential equations and in inverse problems. As shown in the Appendix, the DtN operator is monotone as a function of frequency. This generalization of the monotonicity of 1D impedance is interesting  but unfortunately not sufficient to prove the BBCP in 2D mathematically.
%
%%%%%%%%%%%%%%%%%%%%%%%%%%%%%%%%%%%%%%%%%%%%%%%%%%%%%%%%%%%%%%%%%%%%%
\section{Dirichlet-to-Neumann (DtN) operators: Definition and computation}\label{sec:DtN}
%%%%%%%%%%%%%%%%%%%%%%%%%%%%%%%%%%%%%%%%%%%%%%%%%%%%%%%%%%%%%%%%%%%%%
%
The main stage of the analysis has to do with the behavior of fields in a single semi-infinite strip $S$ composed of repeated lattice cells $C = [0, a] \times [-a/2, a/2]$. The electric field satisfies the generalized Helmholtz equation \eqref{eqn:div-rho-grad-e-k2-eps-e}, the Bloch boundary condition at $\tau = \pm a/2$, and the decay condition \eqref{eqn:e-0-at-infty} at infinity.
The Dirichlet-to-Neumann (DtN) map $\calD$ relates the values of $e$ and $\partial_n e$ at the boundary $n = 0$ of $S$: 
\begin{equation}\label{eqn:DtN-NtD-strip}
	\partial_n e(0, \, \tau; \K, \beta) ~=~ \calD(\K, \beta) \, e(0, \, \tau; \K, \beta)
\end{equation}
Since the $n$-derivative does not affect the complex exponential $\exp (i \beta \tau)$, one can easily alternate between $e$ and $\tilde{e}$ \eqref{eqn:e-eq-tilde-e-exp} in the expressions below.
Recall that the material immediately adjacent to the interface $n=0$ is assumed nonmagnetic; otherwise the magnetic tensor at $n=0$ would appear in the matching conditions. Function arguments will be omitted when there is no risk of confusion.

Our starting point is the computation of the DtN operator for the wave equation \eqref{eqn:div-rho-grad-e-k2-eps-e} in a single lattice cell $C$. The boundary conditions are
\begin{equation}\label{eqn:DtN-cell-Dirichlet-bc}
  \tilde{e}(n, -a/2) = \tilde{e}(n, a/2);
  \quad
  \tilde{e}(0, \tau) = \tilde{f}_0(\tau);
  \quad
  \tilde{e}(a, \tau) = \tilde{f}_a(\tau)
\end{equation}
where $\tilde{f}_0$, $\tilde{f}_a$ are given $\tau$-periodic functions, which can be conveniently represented in the Fourier basis $\psi_m(\tau) = \exp(2\pi i \, m \tau / a)$, $m = 0, \pm 1, \ldots$. This basis is theoretically infinite but can be truncated to a finite one in practical computation. The normal derivatives $\partial_n \tilde{e}$ can also be represented in the same Fourier basis.
Using the subscripts `$l$' and `$r$' as shorthand for the ``left'' ($n = 0$) and ``right'' ($n = a$) edges of the lattice cell $C$, one can write the DtN map (either exact or approximate) in block form:
\begin{equation}\label{eqn:DtN-map-block}
	\partial_n 
	\begin{pmatrix}
		e_{l}\\
		e_{r}
	\end{pmatrix} 
	~=~
	\begin{pmatrix}
	\calD_{ll}^C & \calD_{lr}^C\\
	\calD_{rl}^C & \calD_{rr}^C
\end{pmatrix}
	\begin{pmatrix}
	e_{l}\\
	e_{r}
\end{pmatrix} 
\end{equation}
where the superscript `$C$' refers to the DtN map for the lattice cell.
As an alternative to well-known methods of computing Bloch modes, one can, as first proposed in \cite{Yuan-Lu-DtN-PhC2006,Hu-Lu-DtN-PhC2008}, consider \eqref{eqn:DtN-map-block} in conjunction with the Bloch conditions in the normal direction:
\begin{equation}\label{eqn:eright-eq-lambda-eleft}
    e_{r} = \lambda_B e_{l};
    \quad
    \partial_n e_{r} = \lambda_B \, \partial_n  e_{l};
    	\quad \quad
    \lambda_B = \exp(iqa)
\end{equation}
%
%eq.~\eqref{eqn:DtN-map-block} becomes
%
%\begin{equation}\label{eqn:DtN-map-via-eleft}
%	\partial_n 
%	\begin{pmatrix}
%		e_{l}\\
%		e_{r}
%	\end{pmatrix} 
%	~=~
%	\begin{pmatrix}
%		\calD_{ll}^C + \lambda_B \calD_{lr}^C\\
%		\calD_{rl}^C + \lambda_B \calD_{rr}^C
%	\end{pmatrix} \,
%    e_{l}
%\end{equation}
%
One advantage of the DtN approach is that it generalizes in a natural way to the computation of eigenmodes in a semi-infinite strip and, ultimately, to interface modes between two periodic structures.
Elimination of $\partial_n e_r$ from \eqref{eqn:DtN-map-block}, \eqref{eqn:eright-eq-lambda-eleft} leads to the following quadratic eigenproblem: % for the Bloch parameter $\lambda_B$:
\begin{equation}\label{eqn:DtN-eigenproblem-quadr}
	\left[
	\lambda_B^2 \calD_{lr}^C + \lambda_B(\calD_{ll}^C - \calD_{rr}^C) - \calD_{rl}^C )
	\right]  e_{l} ~=~ 0;
\end{equation}
In \cite{Yuan-Lu-DtN-PhC2006,Hu-Lu-DtN-PhC2008}, analytical solutions for cylindrical inclusions and collocation on the lattice cell boundary are employed. In the present paper, the finite element method (FEM) is used to solve $2(2M+1)$ Dirichlet problems \eqref{eqn:DtN-cell-Dirichlet-bc} in the Fourier basis $\{ \tilde{\psi}_m(\tau) \}$ limited to orders $m = -M, \ldots, M$. One then obtains a set of $2(2M + 1)$ Bloch modes -- propagating for $|\lambda_{B,m}| = 1$ and exponential otherwise. The \textit{decaying} eigenmodes are then used to construct the DtN operator for the semi-infinite strip $S$. This extension of the DtN-Bloch method \cite{Yuan-Lu-DtN-PhC2006,Hu-Lu-DtN-PhC2008} was previously proposed and applied to bound states in the continuum at the surface of a photonic crystal in Ref.~\cite{Hu-Lu-Bound-states2017}, where further details pertaining to this idea can be found.
%Let $\{\underline{e}_{l} \}$ be the set of evanescent solutions of \eqref{eqn:DtN-eigenproblem-quadr}; this set will serve as a basis for expanding the field in the whole strip.
%
The DtN operator is then converted, by a linear transformation, from this evanescent basis to the  Fourier basis $\{ \tilde{\psi}_m(\tau) \}$ along the boundary $n = 0$, $-a/2 \leq \tau \leq a/2$. The numerical procedure is not central to the present paper and is not elaborated upon. 

In 1D, Maxwell's interface conditions between two heterostructures simply amount to impe\-dance (or admittance) matching. A 2D generalization of these conditions is
\begin{equation}\label{eqn:Null-DtN12}
	\text{Null}~ (\DtN_1 + \DtN_2) \neq \varnothing
\end{equation}
The plus sign in the expression is due to the fact that the DtN operator in each structure is defined relative to its own normal direction, \textit{into} the respective domain. The null space, if not empty, contains the boundary value (or values) $e(0, \tau)$ of the possible mode(s) in a chosen basis.
%
%%%%%%%%%%%%%%%%%%%%%%%%%%%%%%%%%%%%%%%%%%%%%%%%%%%%%%%%%%%%%%
\section{Numerical examples}\label{sec:Examples}
%%%%%%%%%%%%%%%%%%%%%%%%%%%%%%%%%%%%%%%%%%%%%%%%%%%%%%%%%%%%%%
%
The purpose of the numerical simulations in this section is to investigate the canonical setup: gyromagnetic structures coupled with simple dielectrics/magnetics \cite{Wang-Chong-Joannopoulos-PRL2008,Wang-Chong-Joannopoulos-Nature2009,Zhao-WeiSha-Chern20}. Due to the presence of gyromagnetic materials such as YiG, the time reversal symmetry is broken, and some of the relevant topological invariants -- the Chern numbers \cite{Wang-Chong-Joannopoulos-PRL2008,Wang-Chong-Joannopoulos-Nature2009,Khanikaev-Shvets-2D-topological17, Vanderbilt-Berry-phases-book2018} --  may be nonzero. 

\begin{figure}
    \centering
    \includegraphics[height=1.9in]{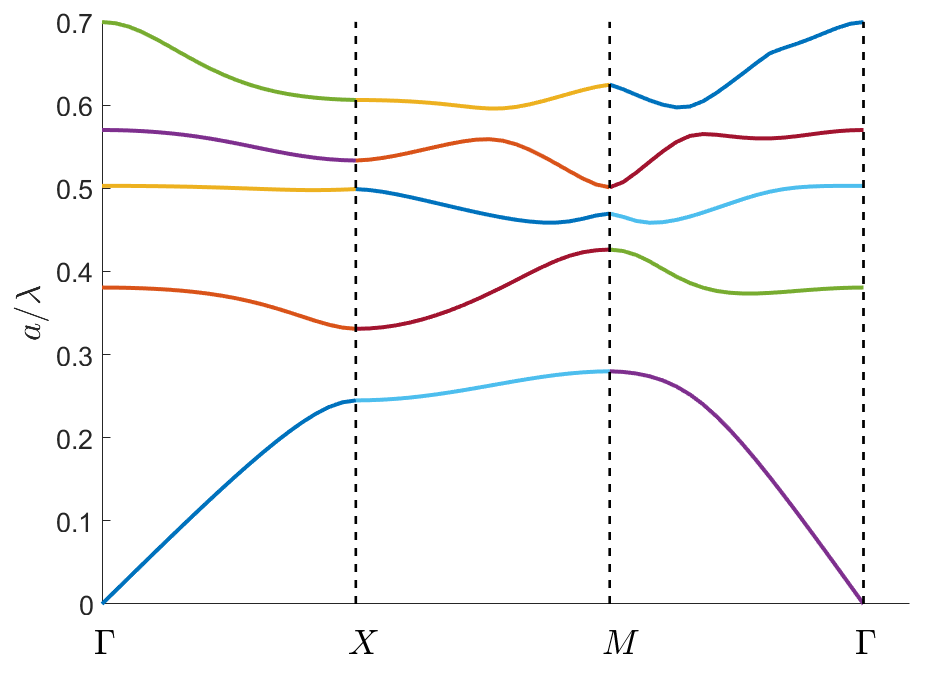}
    \includegraphics[height=1.9in]{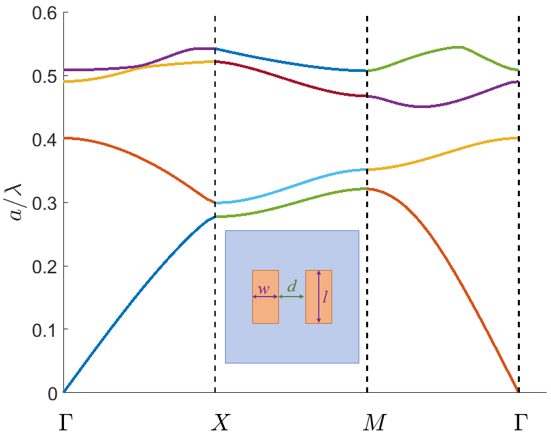}
    \caption{Band diagrams for a gyromagnetic (left) and simple dielectric/magnetic structures (right). 
%    	Left: $\epsilon_{\text{cyl}} = 20$, $\epsilon_{\text{host}} = 2$; $\mu = 14$, $\kappa = 12.4$. Right: $\epsilon_{\text{bar}} = 6$, $\mu_{\text{bar}} = 6$.
%    Inset: The cell of a simple dielectric/magnetic structure contains two rectangular bars with $w = 0.2a$, $l = 0.4a$, $d = 0.2a$.
    Parameters are specified in the text.}
    \label{fig:Bloch-bands-gyro-simple}
\end{figure}

In the simulations reported below, the cell size $a$ of both abutting structures has for convenience been normalized to unity, and the  parameters were adjusted to ensure an overlap between the bandgaps of the two heterostructures under investigation. Namely, the permittivity of the gyromagnetic cylinders in the first periodic structure is $\epsilon_{\text{cyl}} = 20$, their radius is $r_{\text{cyl}} = 0.11a$. The host medium has $\epsilon_{\text{host}} = 2$, and the permeability tensor $\mu$ has, as in \cite{Wang-Chong-Joannopoulos-PRL2008,Wang-Chong-Joannopoulos-Nature2009,Zhao-WeiSha-Chern20}, $\mu_{nn} = \mu_{\tau \tau} = 14$ on the diagonal and $\mu_{n \tau} = -\mu_{\tau n} = 12.4i$ off the diagonal.
%
%\begin{equation}
%	  \mu_{\text{cyl}} ~=~
%	  \begin{pmatrix}
%		      \mu & i \kappa\\
%		      -i \kappa & \mu
%        \end{pmatrix},
%        \quad
%        \mu = 14, ~~~ \kappa = 12.4
%\end{equation}
%
The lattice cell of the second structure contains two rectangular bars $0.2a \times 0.4a$ separated by a $0.2a$ gap (\figref{fig:Bloch-bands-gyro-simple}, inset), in an otherwise empty cell. The physical parameters of the bars are $\epsilon_{\text{bar}} = 6$, $\mu_{\text{bar}} = 6$. Then there exists an overlap in the band gaps of the two structures around the normalized frequency $a / \lambda = \omega a / (2\pi c) \sim 0.42$; see the band diagrams in \figref{fig:Bloch-bands-gyro-simple}.

%\begin{figure}
%	\centering
%   \includegraphics[width=0.25\linewidth]{setup_2bars.png}
%	\caption{The square lattice cell $a \times a$ of a simple dielectric/magnetic structure contains two rectangular bars with $w = 0.2a$, $l = 0.4a$, $d = 0.2a$.}
%	\label{fig:2bars-setup}
%\end{figure} 

The simulations were carried out using FEM with first order triangular elements; the typical number of elements in the lattice cell of each heterostructure was on the order of 5,000, although multiple simulations with finer and coarser meshes were also performed to verify convergence. The DtN map for each of the two structures was computed in the basis of Fourier harmonics as described in \sectref{sec:DtN}. Two main advantages of FEM-DtN are worth noting: (a) applicability to any composition of the lattice cell, and (b) computational efficiency\footnote{The computational cost for the eigenproblem \eqref{eqn:DtN-eigenproblem-quadr} is in practice negligible, since the field at the interface is usually smooth and good accuracy can be attained with a small number of Fourier harmonics. At the same time, sparsity of FE matrices is easier to exploit in Dirichlet solvers than in eigensolvers with large matrices.}.

\begin{figure}
    \centering
    \includegraphics[width=0.48\linewidth]{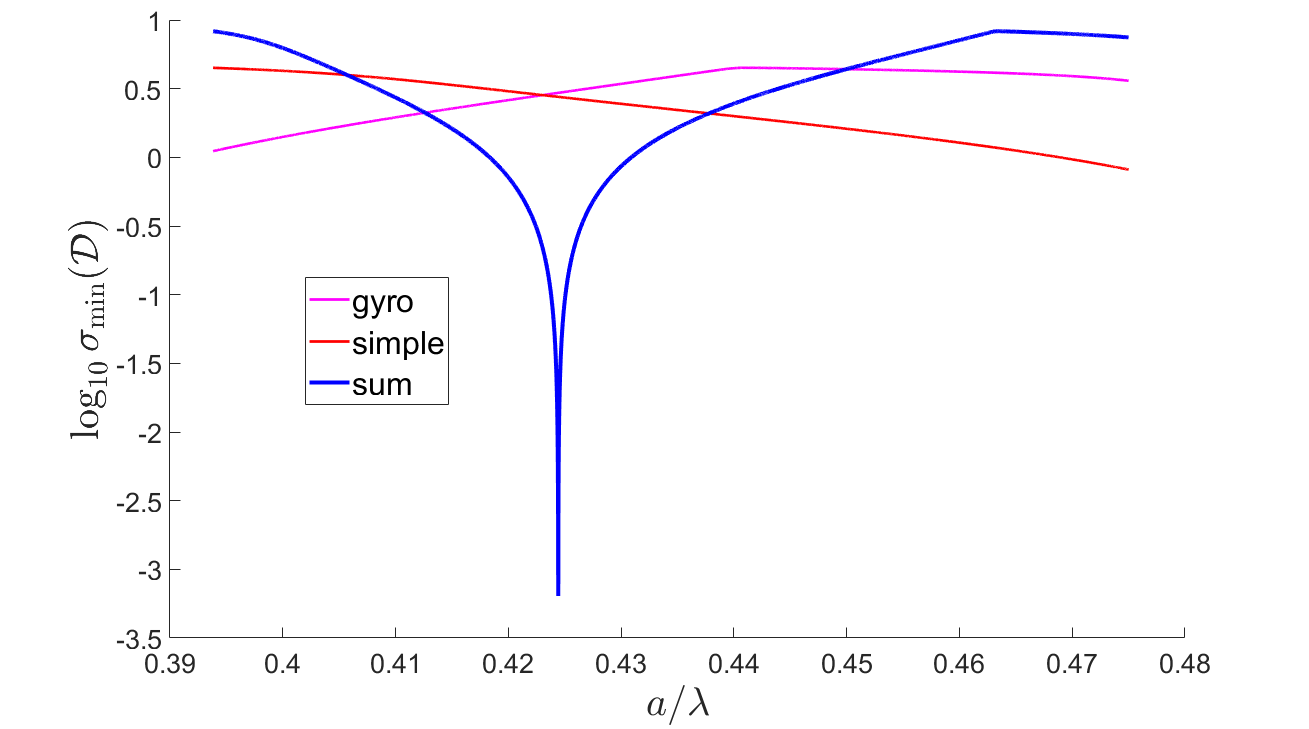}
	\includegraphics[width=0.48\linewidth]{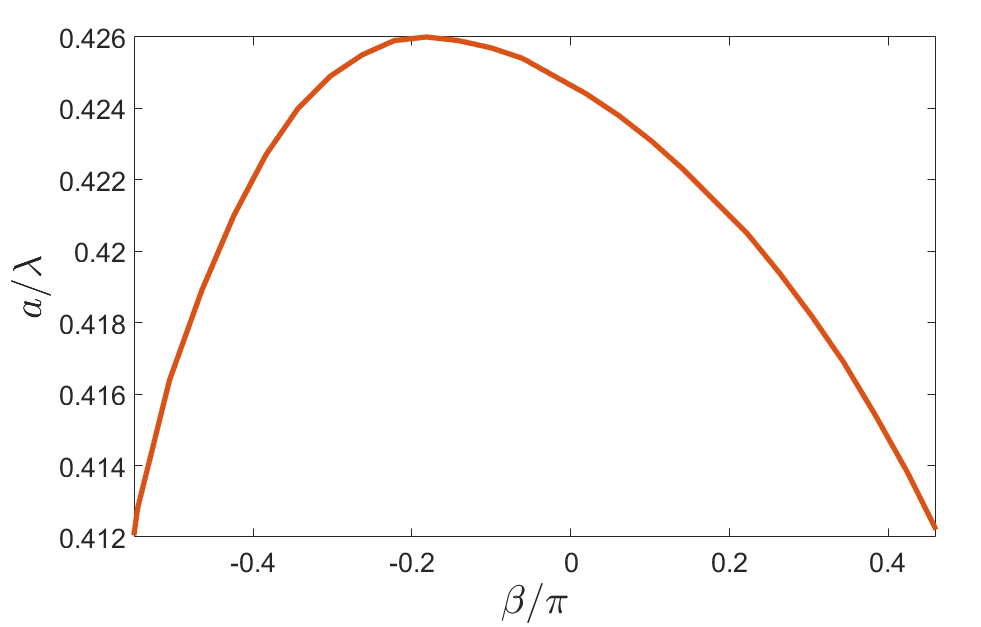}
    \caption{Left: The minimum singular value $\sigma_{\min} (\DtN_1 + \DtN_2)$ exhibits a sharp drop at a particular frequency within a gap, in contrast with the smooth variation of the individual minimum singular values of the two DtN operators. Right: the dispersion curve for the evanescent edge mode. Parameters are specified in the text.
%    	on the gyromagnetic side, $\epsilon_{\text{cyl}} = 20$, $\epsilon_{\text{host}} = 2$, $\mu = 14$, $\kappa = 12.4$, $r_{\text{cyl}} = 0.11a$; on the simple medium side, $\epsilon_{\text{bar}} = 6$, $\mu_{\text{bar}} = 6$; the rectangular bars $0.2a \times 0.4a$ are separated by a $0.2a$ gap.)
    }
    \label{fig:sigmin-emode-gyro-2bars}
\end{figure}

In light of the boundary condition \eqref{eqn:Null-DtN12}, of principal interest is the minimum singular value $\sigma_{\min} (\DtN_1 + \DtN_2)$. \figref{fig:sigmin-emode-gyro-2bars} (left) demonstrates a sharp drop in $\sigma_{\min}$ at a particular frequency ($a / \lambda \sim 0.42$) within the overlap between the two band gaps -- in contrast with the smooth variation of the minimum singular values of the two DtN operators separately. Thus, a qualitative change  occurs when two structures are put together and share a common interface. The maximum order of spatial harmonics is $M = 4$, so that their total number is $2M+1 = 9$. Very similar results were obtained with $M = 3$ and $M = 5$.

%The dispersion curve of the interface mode, shown in \figref{fig:sigmin-emode-gyro-2bars}, right, illustrates an underappreciated point. The difference of the gap Chern numbers $|C_1 - C_2|$ of the respective structures is equal \textit{not} to the total number of boundary modes but, rather, to the ``net current'' (a.k.a. ``spectral flow'' in electronic structure theory) -- the \textit{difference} between the number of modes traveling in one direction and the other \cite[p.~5]{Liu-topological-superconductors-2017}, \cite{Chan-private2024}. The dispersion curve indicates the presence of \textit{two} modes traveling in the opposite directions, implying that the ``net flow'' at the interface is zero; this is consistent with the gap Chern number being zero. Indeed, the Chern numbers of the gyromagnetic structure, calculated as in \cite{Fukui-Chern-numbers05,Zhao-WeiSha-Chern20}, happen to be zero for the first two bands. Hence the Chern number for the second \textit{gap} (around $a / \lambda \sim 0.42$), which by definition is equal to the sum of the two underlying \textit{band} Chern numbers, is also zero.

The dispersion curve of the interface mode, shown in \figref{fig:sigmin-emode-gyro-2bars}, right, highlights an often-overlooked point: the difference in gap Chern numbers $|C_1 - C_2|$ of the two structures is equal \textit{not} to the total number of boundary modes but, rather, to the ``net current'' (a.k.a. ``spectral flow'' in electronic structure theory) -- the \textit{difference} between the number of modes traveling in opposite directions \cite[p.~5]{Liu-topological-superconductors-2017}, \cite{Chan-private2024}. The dispersion curve shows two such modes, resulting in a zero net flow; this aligns with the gap Chern number being zero. Indeed, the Chern numbers for the first two bands of the gyromagnetic structure, calculated using the Fukui method \cite{Fukui-Chern-numbers05,Zhao-WeiSha-Chern20}, are zero. Therefore, the Chern number for the second gap (around $a/\lambda \sim 0.42$) — defined as the sum of the Chern numbers of the underlying bands — is also zero.

\begin{figure}
	\centering
	\includegraphics[width=0.48\linewidth]{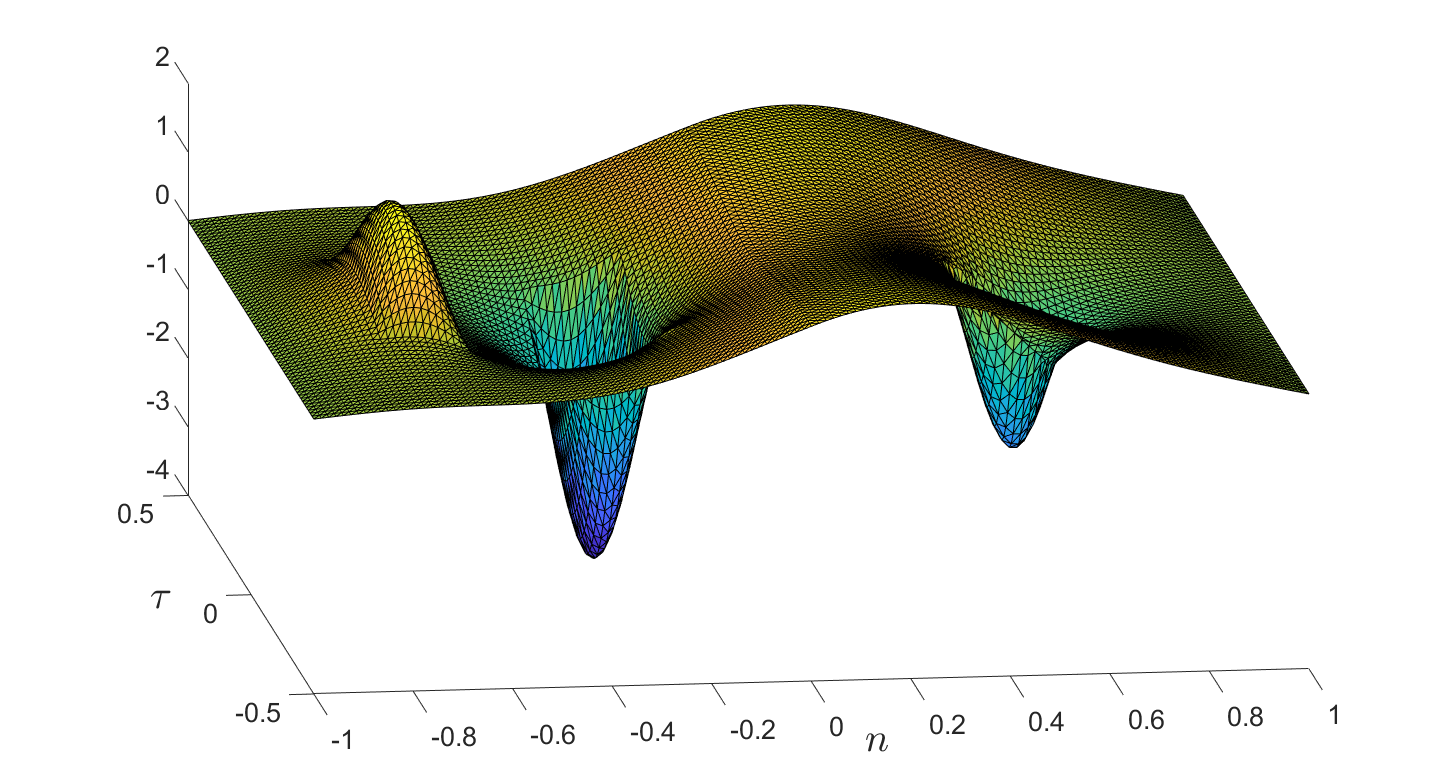}
	\includegraphics[width=0.48\linewidth]{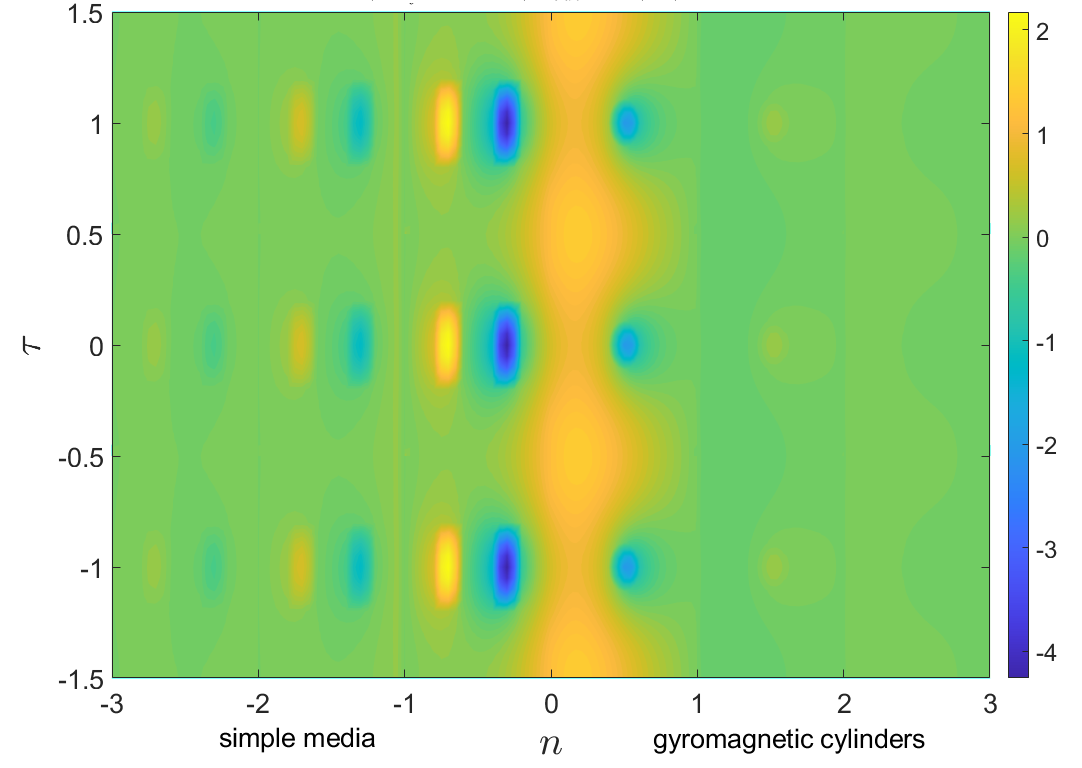}
	\caption{The real part of the evanescent electric field corresponding to $\sigma_{\min} (\DtN_1 + \DtN_2)$ for $\beta = 0$; other parameters are specified in the text. Left: a surface plot for two lattice cells adjacent to the interface. Right: a color plot for multiple cells; a rapid decay of the field away from the interface is manifest.
		%		The simple and gyromagnetic media occupy the half-spaces $n \leq 0$ and $n \geq 0$, respectively.
 }
	\label{fig:Re-efield-gyro-simple-beta0-eps20-mu14-kap124-r011-eps6-mu6-4harm}
\end{figure}

\begin{figure}
	\centering
	\includegraphics[width=0.48\linewidth]{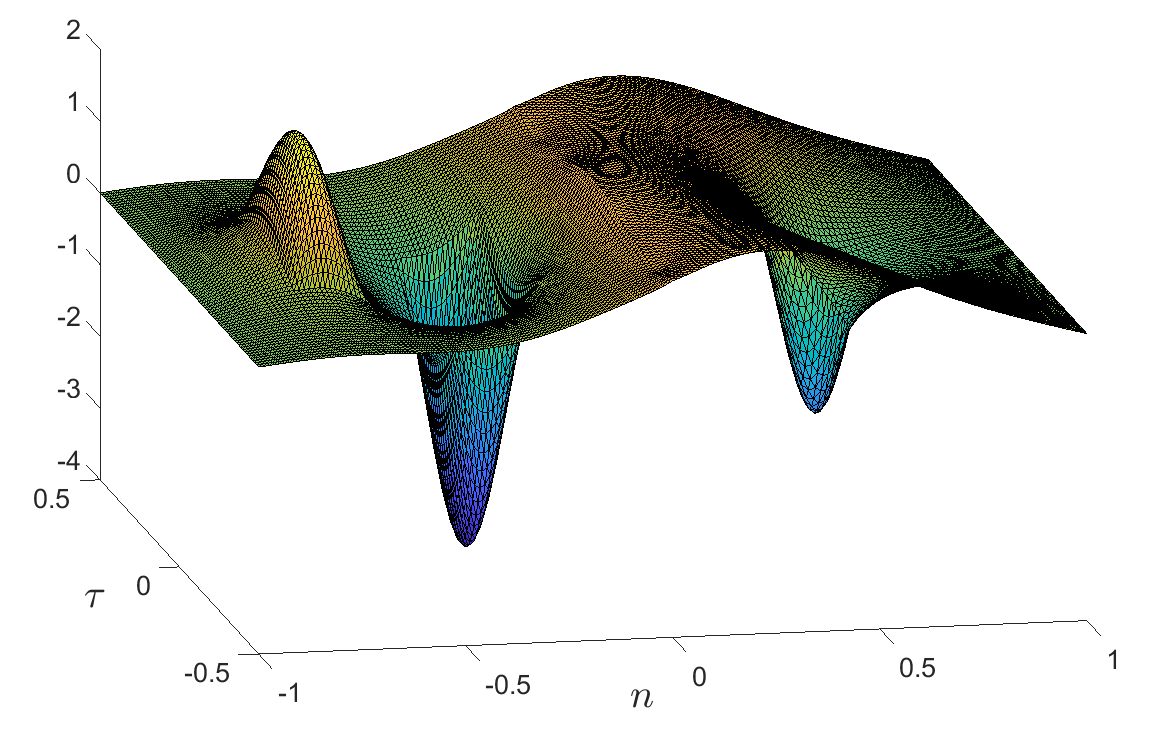}
	\includegraphics[width=0.48\linewidth]{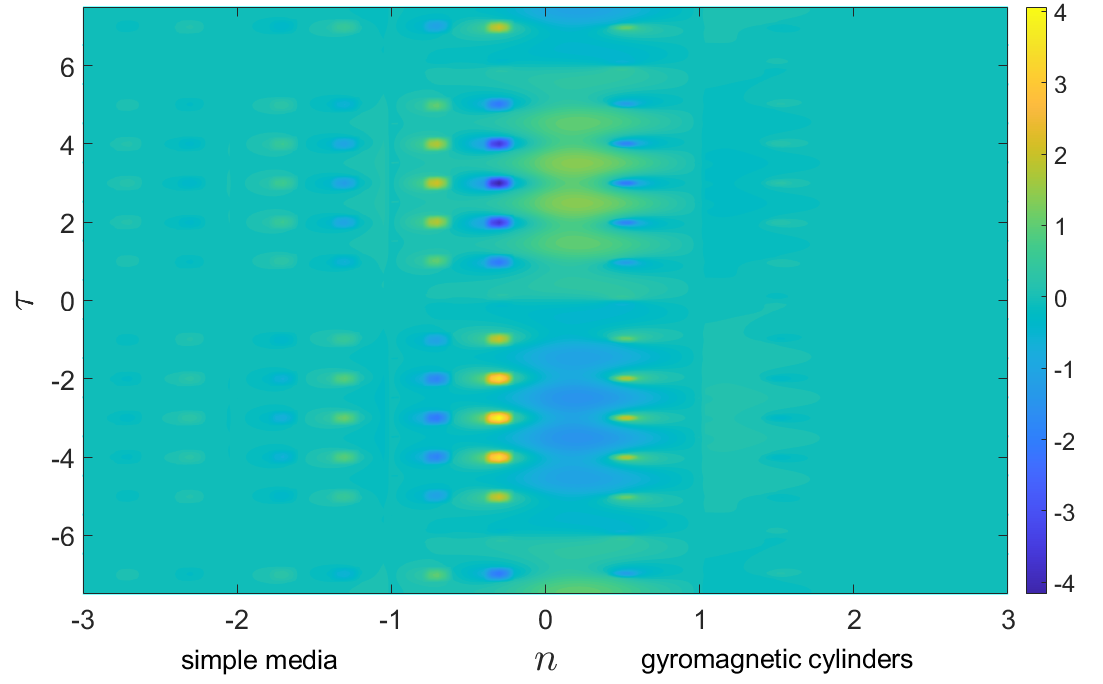}
	\caption{Same as \figref{fig:Re-efield-gyro-simple-beta0-eps20-mu14-kap124-r011-eps6-mu6-4harm}, but for $\beta = \pi/6$.
	}
	\label{fig:Re-e-YiG-simple-alambda-042-beta-pi6-eps20-6-mu14-6-kap124-r011-2bars-epsout-21-h00125-5harm}
\end{figure}

The real part of the electric fields corresponding to $\sigma_{\min} (\DtN_1 + \DtN_2)$ for $\beta a = 0$  and $\beta a = \pi/6$ are illustrated in Figs.~\ref{fig:Re-efield-gyro-simple-beta0-eps20-mu14-kap124-r011-eps6-mu6-4harm} and \ref{fig:Re-e-YiG-simple-alambda-042-beta-pi6-eps20-6-mu14-6-kap124-r011-2bars-epsout-21-h00125-5harm}, respectively. The first field is $\tau$-periodic, while the second one is a wave traveling along the interface. 

In addition to the geometric setup with the two bars, multiple simulations with \textit{cylindrical} inclusions in the dielectric/magnetic structure were performed; parameters were chosen to produce approximately the same bandgap overlap at $a / \lambda \sim 0.42$. The results in all these cases were similar not only qualitatively but quantitatively, and therefore are not reported here.

%%%%%%%%%%%%%%%%%%%%%%%%%%%%%%%%%%%%%%%%%%%%%%%%%%%%%%%%%%%%%%
\section{Discussion and conclusion}\label{sec:Conclusion}
%%%%%%%%%%%%%%%%%%%%%%%%%%%%%%%%%%%%%%%%%%%%%%%%%%%%%%%%%%%%%%
%
The objective of the paper is to examine, analytically and numerically, the validity of the BBCP for problems of 2D Maxwell's electrodynamics. The study is limited to rectangular lattices, although some of the methods and considerations are applicable to hexagonal lattices (``valley photonics'' and the ``photonic spin Hall effect'' \cite{%Ma-Shvets-topological-waveguides15,
	Khanikaev-Shvets-2D-topological17, Xue-Yang-Zhang-Valley-photonics21}). Other topological issues, such as waveguides \cite{Silva-Silveirinha-Tamm-states-2022,Qiu-Zhang-waveguide2023} and non-Hermitian topological photonics are outside the scope of the paper.

The BBCP has been mathematically proven for a range of cases in 1D and 1.5D. The qualitative difference with 2D is that the boundary admittance, a scalar quantity in 1D, turns into an infinite-dimensional DtN map. In the paper, this map is defined and computed for semi-infinite strips in the basis of spatial harmonics along the boundary. A very similar approach was proposed in \cite{Hu-Lu-Bound-states2017} for the computation of bound states in the continuum at a photonic crystal/air interface.

A sufficient and necessary condition for the existence of an interface mode within an overlap of the bandgaps is for the sum of the DtN maps of the two abutting structures to  be singular. This is detected numerically as a sharp minimum of the singular value of the sum of the DtN operators in a basis of spatial harmonics.

From a purely mathematical standpoint, the numerical results cannot be fully conclusive, as the singularity of an infinite-dimensional operator cannot generally be verified exactly -- not to mention that numerical studies are by necessity limited in scope and lack  generality.
But from a physical perspective, results presented in this paper are persuasive in several important respects. First, they give a strong indication that the BBCP is indeed exact in 2D Maxwell's electrodynamics, despite the qualitative differences with 1D and with discrete lattice models. 
Second, the numerical simulations illustrate the fact that the difference in the gap Chern numbers of the two periodic structures is equal \textit{not} to the total number of interface modes but rather to the ``net current'' of these modes.
Third, the analysis suggests a connection between the BBCP and the positivity of electromagnetic energy density.

\section{Acknowledgment}
Communication with Hai Zhang, Che Ting Chan, Didier Felbacq, Ya Yan Lu, Ralf Hiptmair and Vadim Markel is very gratefully acknowledged.

%%%%%%%%%%%%%%%%%%%%%%%%%%%%%%%%%%%%%%%%%%%%%%%%%%%%%%%%%%%%%%%%%%
\section*{Appendix: Monotonicity of the DtN operator in a gap}\label{sec:Monotonicity-DtN-in-gap}
%%%%%%%%%%%%%%%%%%%%%%%%%%%%%%%%%%%%%%%%%%%%%%%%%%%%%%%%%%%%%%%%%%
%
The 2D analysis consists of two main steps: 1) differentiate the wave equation \eqref{eqn:div-rho-grad-e-k2-eps-e} with respect to $s = \K^2 = (\omega / c)^2$, leading to an equation for the derivative $g = \partial_s e$; 2) inner-multiply this equation for $g$ with $e$ and integrate by parts, arriving at the boundary values of the fields and at a relationship for the DtN operator. For 1D problems, this plan was pioneered in \cite{Thiang-Zhang-Bulk-interface-1D-2023} and slightly simplified in \cite{Tsukerman-Markel-Topology-Bloch-impedance2024}. % Below are the technical details.
In 1D, the boundary admittance (inverse of impedance) is a scalar; in 2D, it turns into the Dirichlet-to-Neumann (DtN) operator $\DtN(s)$ (for brevity, dependence on $\beta$ is not explicitly indicated):
\begin{equation}\label{eqn:dne-eq-De}
	\partial_n e(0, \tau; s)  ~=~ \DtN(s) \,  e(0, \tau; s) 
\end{equation}
This operator is self-adjoint in the space of functions satisfying the Bloch conditions at $\tau = \pm a/2$ and decaying at $n \rightarrow \infty$. This can be shown by the standard integration-by-parts argument, keeping in mind that the functions are complex valued. Indeed, let $u = u(n, \tau; s)$, $v = v(n, \tau; s)$ be two arbitrary solutions of 
problem \eqref{eqn:Maxwell}, \eqref{eqn:Bloch-conditions-tau}, \eqref{eqn:bc-infinity} in $S$. The arguments of the functions will be dropped whenever this does not cause confusion; then
\begin{equation}\label{eqn:Greens-formula-u}
	0 = (\nabla \cdot (\rho \nabla u) + s \epsilon u, \, v )  
	\, \overset{\text{Green's~id.}}{=} \,
	-[\partial_n u, \, v]  -
	(\rho \nabla u, \, \nabla v) + s (\epsilon u, \, v ) 
\end{equation}
\begin{equation}\label{eqn:Greens-formula-v}
   0 = (\nabla \cdot (\rho \nabla v) + s \epsilon v, \, u)  
   \, \overset{\text{Green's~id.}}{=} \,
   -[\partial_n v, \, u]  -
   (\rho \nabla v, \, \nabla u) + s (\epsilon v, \, u ) 
\end{equation}
The inner products over the strip $S$ and its boundary $n=0$ are defined as follows:
\begin{equation}\label{eqn:inner-product-S-n0}
	(u, \, v )  = \int_S uv^* \, dS;
%\end{equation}
%%
%\begin{equation}\label{eqn:inner-product-n0}
	\quad
	[u, \, v ]  = \int_{-a/2}^{a/2} u(0, \tau) \, v^*(0, \tau) \, d \tau
\end{equation}
The negative sign of the boundary terms in \eqref{eqn:Greens-formula-u}, \eqref{eqn:Greens-formula-v} reflect the fact that the $n$ coordinate is directed \textit{into} the domain.
In the process of integration by parts, the contribution at $n \rightarrow \infty$ vanishes, and the boundary terms at $\tau = \pm a/2$ cancel out due to the conjugate exponentials $\exp (\pm i \beta a/2)$.
Noting that $\epsilon$ is real and $\rho$ is Hermitian by our assumptions, we obtain after subtracting  \eqref{eqn:Greens-formula-u} from the conjugate of \eqref{eqn:Greens-formula-v}:
\begin{equation}\label{eqn:dnu-v-minus-u-dnv-eq-0}
	[\partial_n u, \, v]  - [u, \partial_n v] = 0
	\quad \Leftrightarrow \quad
%\end{equation}
%
%Since $\partial_n v = \DtN v$ and $\partial_n u = \DtN u$, the self-adjointness of $\DtN$ is manifest:
%%
%\begin{equation}\label{eqn:Duv-eq-u-Dv}
	[\DtN u, \, v]  ~=~ [u, \DtN v]
\end{equation}
We now proceed to the main part of the analysis. The governing equation \eqref{eqn:div-rho-grad-e-k2-eps-e} for $e$, with $s = \K^2$, is
\begin{equation}\label{eqn:div-rho-grad-e-s-eps-e}
%	\nabla \cdot \left( \rho(n, \tau) \nabla  e(n, \tau; s) \right) + s \epsilon(n, \tau; s) \, e(n, \tau; s) 	
    \nabla \cdot \left( \rho \nabla  e \right) ~+~ s \epsilon e ~=~ 0
\end{equation}
%
%Dependence on $\beta$ is not explicitly shown to make the notation less cumbersome. 
%
Differentiation of \eqref{eqn:div-rho-grad-e-s-eps-e} and of \eqref{eqn:dne-eq-De} with respect to $s$ yields, respectively,
\begin{equation}\label{eqn:nabla-rho-nabla-we-seps}
%	\nabla \cdot (\rho(n, \tau) \nabla g(n, \tau; s)) + w(n, \tau; s) \, e(n, \tau; s) +
%	s \epsilon(n, \tau; s) \, g(n, \tau; s)  = 0
	\nabla \cdot (\rho \nabla g) \,+\, w e \,+\, s \epsilon g  = 0
\end{equation}
\begin{equation}\label{eqn:dnq-eq-Dprime-e-Dg}
	\partial_n g(0, \tau; s)  ~=~ \left[ \partial_s \DtN(s) \right] e(0, \tau; s) \,+\, \DtN(s) \,  g(0, \tau; s) 
\end{equation}
where the following notation for the $s$-derivatives has been used:
\begin{equation}\label{eqn:g-w-define}
	g(n, \tau; s) = \partial_s e(n, \tau; s);
	\quad
	w(n, \tau; s) = \partial_s \left[ s \epsilon(n, \tau; s) \right]
\end{equation}
Inner-multiplying \eqref{eqn:nabla-rho-nabla-we-seps} with $e$ we get, after complex conjugation, and keeping in mind that $\epsilon$ and $w$ are real under our assumptions,
\begin{equation}\label{eqn:nabla-rho-nabla-ge-etc}
	(\nabla \cdot (\rho \nabla g), \, e)^* + (w e, \, e) +  s (\epsilon e, \, g) = 0	
\end{equation}
\noindent
 The last term can be transformed due to \eqref{eqn:div-rho-grad-e-s-eps-e}, and we have
\begin{equation}\label{eqn:nabla-rho-nabla-gestar-etc}
   (\nabla \cdot (\rho \nabla g), \, e)^* ~+~ (w e, \, e)
              ~-~ \left( \nabla \cdot \left( \rho \nabla  e \right), \, g \right) = 0
\end{equation}
%
%where we used the assumption that $\epsilon$ is real and $g = g^*$ is Hermitian.
Applying Green's identity to the two divergence terms, recalling that $\rho(0, \tau) = 1$ by assumption and $n$ is directed inward, one obtains
\begin{equation}\label{eqn:dne-g-e-dng-w-ee}
	\left[ \partial_n e, \, g \right] ~-~ \left[e, \, \partial_n g \right] ~+~ (w e, e) ~=~ 0
\end{equation}
Note parenthetically that in general $\partial_n g \neq \DtN g$, since $g$ is not a solution of the wave equation; thus, the inner products in \eqref{eqn:dne-g-e-dng-w-ee} do not cancel despite the Hermiticity of $\DtN$. 
Of the boundary terms, only the $n = 0$ integrals have survived in \eqref{eqn:dne-g-e-dng-w-ee}; all other contributions vanish due to the Bloch periodicity of functions in the $\tau$ direction and their decay at infinity. 

We now substitute into \eqref{eqn:dne-g-e-dng-w-ee} expressions \eqref{eqn:dne-eq-De} for $\partial_n e(0, \tau; s)$ and \eqref{eqn:dnq-eq-Dprime-e-Dg} for $\partial_n g(0, \tau; s)$:
\begin{equation}\label{eqn:dne-g-minus-Dprime-e-etc}
	\left[ \DtN e, \, g \right] ~-~ \left[e, \, (\partial_s \DtN) \,  e + \DtN \,  g \right]
	~+~ (w e, e) ~=~ 0
\end{equation}
The two terms containing $\DtN$ in \eqref{eqn:dne-g-minus-Dprime-e-etc} cancel out because the DtN operator is self-adjoint, and this equation simplifies:
\begin{equation}\label{eqn:dn-e0-g0-int-w-e2-simplified}
	 	\left[(\partial_s \DtN)  e, \,  e \right] ~=~ (w e, e)
\end{equation}
The right hand side of this relation is physically significant: the quantity $w(\omega) = \partial_s (s \epsilon(s)) ~
= \epsilon(\omega) + \frac12 \omega \, \partial_{\omega} \epsilon(\omega)$ is closely related to the Brillouin-Landau-Lifshitz definition of electromagnetic energy density $w_{\text{BLL}} = \partial_{\omega} (\omega \epsilon(\omega)) = \epsilon(\omega) + \omega \partial_{\omega} \epsilon(\omega)$ 
	\cite[p. 92]{Brillouin60},  \cite[\S 80, \S 84]{Landau84}, which must be positive in any physical medium;
%%
%\begin{equation}\label{eqn:w-BLL}
%	w_{\text{BLL}} = \partial_{\omega} (\omega \, \epsilon(\omega))
%	= \epsilon(\omega) + \omega \, \partial_{\omega} \epsilon(\omega)
%\end{equation}
%%
%while, from \eqref{eqn:g-w-define},
%%
%\begin{equation}\label{eqn:w-ours}
%	w(\omega) = \partial_s (s \epsilon(s)) ~
%	\overset{s \, = \, \omega^2 / c^2}{=} 
%	\epsilon(\omega) + \frac12 \, \omega \, \partial_{\omega} \epsilon(\omega)
%\end{equation}
%
see \cite{Tsukerman-Markel-Topology-Bloch-impedance2024} for a few additional algebraic details.

Since the boundary trace $e(0, \tau; s)$ in \eqref{eqn:dn-e0-g0-int-w-e2-simplified} can be arbitrary, we have arrived at the following generalization of the monotonicity of impedance. For any physically realizable structures (with a positive density of electromagnetic energy), the $s$-derivative of the DtN operator is positive definite. It follows that $\DtN$ is monotonically increasing as an operator in the following sense:
\begin{equation}\label{eqn:Monotonicity-NtD}
	\forall s_1 < s_2, ~~ \DtN(s_1) < \DtN(s_2)
	~~ \Leftrightarrow ~~
	\left( \, \DtN(s_1) \, x, ~ x \, \right) < \left( \, \DtN(s_2) \, x, ~ x \, \right), 
	\quad \forall x \in \text{dom} \, \DtN,  ~~ x \neq 0
\end{equation}
This is true because %$\partial_s \DtN(s)$ is proven to be positive definite, and sums or 
integrals of positive definite operators are themselves positive definite:
\begin{equation}
	\DtN(s_2) ~=~ \DtN(s_1) + \int_{s_1}^{s_2} \partial_s \DtN(s) \, ds
	~>~ \DtN(s_1) \quad \quad \therefore
\end{equation}

%%%%%%%%%% If using BibTeX:

%\bibliographystyle{opticajnl}
%\bibliographystyle{plain}
%\bibliographystyle{apsrev4-2}
\bibliography{Topological_photonics}

%apsrev4-2.bst 2019-01-14 (MD) hand-edited version of apsrev4-1.bst
%Control: key (0)
%Control: author (8) initials jnrlst
%Control: editor formatted (1) identically to author
%Control: production of article title (0) allowed
%Control: page (0) single
%Control: year (1) truncated
%Control: production of eprint (0) enabled
\begin{thebibliography}{36}%
\makeatletter
\providecommand \@ifxundefined [1]{%
 \@ifx{#1\undefined}
}%
\providecommand \@ifnum [1]{%
 \ifnum #1\expandafter \@firstoftwo
 \else \expandafter \@secondoftwo
 \fi
}%
\providecommand \@ifx [1]{%
 \ifx #1\expandafter \@firstoftwo
 \else \expandafter \@secondoftwo
 \fi
}%
\providecommand \natexlab [1]{#1}%
\providecommand \enquote  [1]{``#1''}%
\providecommand \bibnamefont  [1]{#1}%
\providecommand \bibfnamefont [1]{#1}%
\providecommand \citenamefont [1]{#1}%
\providecommand \href@noop [0]{\@secondoftwo}%
\providecommand \href [0]{\begingroup \@sanitize@url \@href}%
\providecommand \@href[1]{\@@startlink{#1}\@@href}%
\providecommand \@@href[1]{\endgroup#1\@@endlink}%
\providecommand \@sanitize@url [0]{\catcode `\\12\catcode `\$12\catcode
  `\&12\catcode `\#12\catcode `\^12\catcode `\_12\catcode `\%12\relax}%
\providecommand \@@startlink[1]{}%
\providecommand \@@endlink[0]{}%
\providecommand \url  [0]{\begingroup\@sanitize@url \@url }%
\providecommand \@url [1]{\endgroup\@href {#1}{\urlprefix }}%
\providecommand \urlprefix  [0]{URL }%
\providecommand \Eprint [0]{\href }%
\providecommand \doibase [0]{https://doi.org/}%
\providecommand \selectlanguage [0]{\@gobble}%
\providecommand \bibinfo  [0]{\@secondoftwo}%
\providecommand \bibfield  [0]{\@secondoftwo}%
\providecommand \translation [1]{[#1]}%
\providecommand \BibitemOpen [0]{}%
\providecommand \bibitemStop [0]{}%
\providecommand \bibitemNoStop [0]{.\EOS\space}%
\providecommand \EOS [0]{\spacefactor3000\relax}%
\providecommand \BibitemShut  [1]{\csname bibitem#1\endcsname}%
\let\auto@bib@innerbib\@empty
%</preamble>
\bibitem [{\citenamefont {Haldane}\ and\ \citenamefont
  {Raghu}(2008)}]{Haldane-Raghu-PRL08}%
  \BibitemOpen
  \bibfield  {author} {\bibinfo {author} {\bibfnamefont {F.~D.~M.}\
  \bibnamefont {Haldane}}\ and\ \bibinfo {author} {\bibfnamefont
  {S.}~\bibnamefont {Raghu}},\ }\bibfield  {title} {\bibinfo {title} {Possible
  realization of directional optical waveguides in photonic crystals with
  broken time-reversal symmetry},\ }\href@noop {} {\bibfield  {journal}
  {\bibinfo  {journal} {Phys. Rev. Lett.}\ }\textbf {\bibinfo {volume} {100}},\
  \bibinfo {pages} {013904} (\bibinfo {year} {2008})}\BibitemShut {NoStop}%
\bibitem [{\citenamefont {Wang}\ \emph {et~al.}(2008)\citenamefont {Wang},
  \citenamefont {Chong}, \citenamefont {Joannopoulos},\ and\ \citenamefont
  {Solja\v{c}i\'{c}}}]{Wang-Chong-Joannopoulos-PRL2008}%
  \BibitemOpen
  \bibfield  {author} {\bibinfo {author} {\bibfnamefont {Z.}~\bibnamefont
  {Wang}}, \bibinfo {author} {\bibfnamefont {Y.~D.}\ \bibnamefont {Chong}},
  \bibinfo {author} {\bibfnamefont {J.~D.}\ \bibnamefont {Joannopoulos}},\ and\
  \bibinfo {author} {\bibfnamefont {M.}~\bibnamefont {Solja\v{c}i\'{c}}},\
  }\bibfield  {title} {\bibinfo {title} {Reflection-free one-way edge modes in
  a gyromagnetic photonic crystal},\ }\href@noop {} {\bibfield  {journal}
  {\bibinfo  {journal} {Phys. Rev. Lett.}\ }\textbf {\bibinfo {volume} {100}},\
  \bibinfo {pages} {013905} (\bibinfo {year} {2008})}\BibitemShut {NoStop}%
\bibitem [{\citenamefont
  {Vanderbilt}(2018)}]{Vanderbilt-Berry-phases-book2018}%
  \BibitemOpen
  \bibfield  {author} {\bibinfo {author} {\bibfnamefont {D.}~\bibnamefont
  {Vanderbilt}},\ }\href@noop {} {\emph {\bibinfo {title} {Berry Phases in
  Electronic Structure Theory: Electric Polarization, Orbital Magnetization and
  Topological Insulators}}}\ (\bibinfo  {publisher} {Cambridge University
  Press},\ \bibinfo {year} {2018})\BibitemShut {NoStop}%
\bibitem [{\citenamefont {Banerjee}\ \emph {et~al.}(2023)\citenamefont
  {Banerjee}, \citenamefont {Sarkar}, \citenamefont {Dey},\ and\ \citenamefont
  {Narayan}}]{Banerjee-Non-Hermitian2023}%
  \BibitemOpen
  \bibfield  {author} {\bibinfo {author} {\bibfnamefont {A.}~\bibnamefont
  {Banerjee}}, \bibinfo {author} {\bibfnamefont {R.}~\bibnamefont {Sarkar}},
  \bibinfo {author} {\bibfnamefont {S.}~\bibnamefont {Dey}},\ and\ \bibinfo
  {author} {\bibfnamefont {A.}~\bibnamefont {Narayan}},\ }\bibfield  {title}
  {\bibinfo {title} {{Non-Hermitian topological phases: principles and
  prospects}},\ }\href@noop {} {\bibfield  {journal} {\bibinfo  {journal}
  {Journal of Physics: Condensed Matter}\ }\textbf {\bibinfo {volume} {35}},\
  \bibinfo {pages} {333001} (\bibinfo {year} {2023})}\BibitemShut {NoStop}%
\bibitem [{\citenamefont {Yan}\ \emph {et~al.}(2023)\citenamefont {Yan},
  \citenamefont {Zhao}, \citenamefont {Zhou}, \citenamefont {Ma}, \citenamefont
  {Lyu}, \citenamefont {Chu}, \citenamefont {Hu},\ and\ \citenamefont
  {Gong}}]{Yan-Gong-NonHermitian-topo-review}%
  \BibitemOpen
  \bibfield  {author} {\bibinfo {author} {\bibfnamefont {Q.}~\bibnamefont
  {Yan}}, \bibinfo {author} {\bibfnamefont {B.}~\bibnamefont {Zhao}}, \bibinfo
  {author} {\bibfnamefont {R.}~\bibnamefont {Zhou}}, \bibinfo {author}
  {\bibfnamefont {R.}~\bibnamefont {Ma}}, \bibinfo {author} {\bibfnamefont
  {Q.}~\bibnamefont {Lyu}}, \bibinfo {author} {\bibfnamefont {S.}~\bibnamefont
  {Chu}}, \bibinfo {author} {\bibfnamefont {X.}~\bibnamefont {Hu}},\ and\
  \bibinfo {author} {\bibfnamefont {Q.}~\bibnamefont {Gong}},\ }\bibfield
  {title} {\bibinfo {title} {{Advances and applications on non-Hermitian
  topological photonics}},\ }\href@noop {} {\bibfield  {journal} {\bibinfo
  {journal} {Nanophotonics}\ }\textbf {\bibinfo {volume} {12}},\ \bibinfo
  {pages} {2247} (\bibinfo {year} {2023})}\BibitemShut {NoStop}%
\bibitem [{\citenamefont {Wang}\ \emph {et~al.}(2009)\citenamefont {Wang},
  \citenamefont {Chong}, \citenamefont {Joannopoulos},\ and\ \citenamefont
  {Solja\v{c}i\'{c}}}]{Wang-Chong-Joannopoulos-Nature2009}%
  \BibitemOpen
  \bibfield  {author} {\bibinfo {author} {\bibfnamefont {Z.}~\bibnamefont
  {Wang}}, \bibinfo {author} {\bibfnamefont {Y.}~\bibnamefont {Chong}},
  \bibinfo {author} {\bibfnamefont {J.}~\bibnamefont {Joannopoulos}},\ and\
  \bibinfo {author} {\bibfnamefont {M.}~\bibnamefont {Solja\v{c}i\'{c}}},\
  }\bibfield  {title} {\bibinfo {title} {Observation of unidirectional
  backscattering-immune topological electromagnetic states},\ }\href@noop {}
  {\bibfield  {journal} {\bibinfo  {journal} {Nature}\ }\textbf {\bibinfo
  {volume} {461}},\ \bibinfo {pages} {772} (\bibinfo {year}
  {2009})}\BibitemShut {NoStop}%
\bibitem [{\citenamefont {Khanikaev}\ and\ \citenamefont
  {Shvets}(2017)}]{Khanikaev-Shvets-2D-topological17}%
  \BibitemOpen
  \bibfield  {author} {\bibinfo {author} {\bibfnamefont {A.~B.}\ \bibnamefont
  {Khanikaev}}\ and\ \bibinfo {author} {\bibfnamefont {G.}~\bibnamefont
  {Shvets}},\ }\bibfield  {title} {\bibinfo {title} {Two-dimensional
  topological photonics},\ }\href@noop {} {\bibfield  {journal} {\bibinfo
  {journal} {Nature Photonics}\ }\textbf {\bibinfo {volume} {11}},\ \bibinfo
  {pages} {763} (\bibinfo {year} {2017})}\BibitemShut {NoStop}%
\bibitem [{\citenamefont {Xue}\ \emph {et~al.}(2021)\citenamefont {Xue},
  \citenamefont {Yang},\ and\ \citenamefont
  {Zhang}}]{Xue-Yang-Zhang-Valley-photonics21}%
  \BibitemOpen
  \bibfield  {author} {\bibinfo {author} {\bibfnamefont {H.}~\bibnamefont
  {Xue}}, \bibinfo {author} {\bibfnamefont {Y.}~\bibnamefont {Yang}},\ and\
  \bibinfo {author} {\bibfnamefont {B.}~\bibnamefont {Zhang}},\ }\bibfield
  {title} {\bibinfo {title} {Topological valley photonics: Physics and device
  applications},\ }\href@noop {} {\bibfield  {journal} {\bibinfo  {journal}
  {Advanced Photonics Research}\ }\textbf {\bibinfo {volume} {2}},\ \bibinfo
  {pages} {2100013} (\bibinfo {year} {2021})}\BibitemShut {NoStop}%
\bibitem [{Note1()}]{Note1}%
  \BibitemOpen
  \bibinfo {note} {More precisely, within a frequency overlap of the bandgaps
  of the two adjacent heterostructures.}\BibitemShut {Stop}%
\bibitem [{\citenamefont {Pancharatnam}(1956)}]{Pancharatnam56}%
  \BibitemOpen
  \bibfield  {author} {\bibinfo {author} {\bibfnamefont {S.}~\bibnamefont
  {Pancharatnam}},\ }\bibfield  {title} {\bibinfo {title} {Generalized theory
  of interference, and its applications. {Part I. Coherent} pencils},\
  }\href@noop {} {\bibfield  {journal} {\bibinfo  {journal} {Proc. Indian Acad.
  Sci. A.}\ }\textbf {\bibinfo {volume} {44}},\ \bibinfo {pages} {247}
  (\bibinfo {year} {1956})}\BibitemShut {NoStop}%
\bibitem [{\citenamefont {Klitzing}\ \emph {et~al.}(1980)\citenamefont
  {Klitzing}, \citenamefont {Dorda},\ and\ \citenamefont
  {Pepper}}]{vonKlitzing-QH-effect1980}%
  \BibitemOpen
  \bibfield  {author} {\bibinfo {author} {\bibfnamefont {K.~v.}\ \bibnamefont
  {Klitzing}}, \bibinfo {author} {\bibfnamefont {G.}~\bibnamefont {Dorda}},\
  and\ \bibinfo {author} {\bibfnamefont {M.}~\bibnamefont {Pepper}},\
  }\bibfield  {title} {\bibinfo {title} {New method for high-accuracy
  determination of the fine-structure constant based on quantized hall
  resistance},\ }\href@noop {} {\bibfield  {journal} {\bibinfo  {journal}
  {Phys. Rev. Lett.}\ }\textbf {\bibinfo {volume} {45}},\ \bibinfo {pages}
  {494} (\bibinfo {year} {1980})}\BibitemShut {NoStop}%
\bibitem [{\citenamefont {Hasan}\ and\ \citenamefont
  {Kane}(2010)}]{Hasan-Kane-Colloquium-TI2010}%
  \BibitemOpen
  \bibfield  {author} {\bibinfo {author} {\bibfnamefont {M.~Z.}\ \bibnamefont
  {Hasan}}\ and\ \bibinfo {author} {\bibfnamefont {C.~L.}\ \bibnamefont
  {Kane}},\ }\bibfield  {title} {\bibinfo {title} {Colloquium: Topological
  insulators},\ }\href@noop {} {\bibfield  {journal} {\bibinfo  {journal} {Rev.
  Mod. Phys.}\ }\textbf {\bibinfo {volume} {82}},\ \bibinfo {pages} {3045}
  (\bibinfo {year} {2010})}\BibitemShut {NoStop}%
\bibitem [{\citenamefont {Hatsugai}(1993{\natexlab{a}})}]{Hatsugai-PRL1993}%
  \BibitemOpen
  \bibfield  {author} {\bibinfo {author} {\bibfnamefont {Y.}~\bibnamefont
  {Hatsugai}},\ }\bibfield  {title} {\bibinfo {title} {Chern number and edge
  states in the integer quantum {Hall} effect},\ }\href@noop {} {\bibfield
  {journal} {\bibinfo  {journal} {Phys. Rev. Lett.}\ }\textbf {\bibinfo
  {volume} {71}},\ \bibinfo {pages} {3697} (\bibinfo {year}
  {1993}{\natexlab{a}})}\BibitemShut {NoStop}%
\bibitem [{\citenamefont {Hatsugai}(1993{\natexlab{b}})}]{Hatsugai-PRB1993}%
  \BibitemOpen
  \bibfield  {author} {\bibinfo {author} {\bibfnamefont {Y.}~\bibnamefont
  {Hatsugai}},\ }\bibfield  {title} {\bibinfo {title} {Edge states in the
  integer quantum {Hall} effect and the {Riemann} surface of the {Bloch}
  function},\ }\href@noop {} {\bibfield  {journal} {\bibinfo  {journal} {Phys.
  Rev. B}\ }\textbf {\bibinfo {volume} {48}},\ \bibinfo {pages} {11851}
  (\bibinfo {year} {1993}{\natexlab{b}})}\BibitemShut {NoStop}%
\bibitem [{\citenamefont {Prodan}\ and\ \citenamefont
  {Schulz-Baldes}(2016)}]{Prodan-Schulz-Baldes-2016}%
  \BibitemOpen
  \bibfield  {author} {\bibinfo {author} {\bibfnamefont {E.}~\bibnamefont
  {Prodan}}\ and\ \bibinfo {author} {\bibfnamefont {H.}~\bibnamefont
  {Schulz-Baldes}},\ }\href@noop {} {\emph {\bibinfo {title} {Bulk and Boundary
  Invariants for Complex Topological Insulators}}}\ (\bibinfo  {publisher}
  {Springer International Publishing, Switzerland},\ \bibinfo {year}
  {2016})\BibitemShut {NoStop}%
\bibitem [{\citenamefont {Elbau}\ and\ \citenamefont
  {Graf}(2002)}]{Elbau-Graf-Bulk-edge-conductance-2002}%
  \BibitemOpen
  \bibfield  {author} {\bibinfo {author} {\bibfnamefont {P.}~\bibnamefont
  {Elbau}}\ and\ \bibinfo {author} {\bibfnamefont {G.~M.}\ \bibnamefont
  {Graf}},\ }\bibfield  {title} {\bibinfo {title} {Equality of bulk and edge
  {Hall} conductance revisited},\ }\href@noop {} {\bibfield  {journal}
  {\bibinfo  {journal} {Communications in Mathematical Physics}\ }\textbf
  {\bibinfo {volume} {229}},\ \bibinfo {pages} {415} (\bibinfo {year}
  {2002})}\BibitemShut {NoStop}%
\bibitem [{\citenamefont {Kellendonk}\ and\ \citenamefont
  {Schulz-Baldes}(2004)}]{Kellendonk-Schulz-Baldes-edge-currents-2004}%
  \BibitemOpen
  \bibfield  {author} {\bibinfo {author} {\bibfnamefont {J.}~\bibnamefont
  {Kellendonk}}\ and\ \bibinfo {author} {\bibfnamefont {H.}~\bibnamefont
  {Schulz-Baldes}},\ }\bibfield  {title} {\bibinfo {title} {Quantization of
  edge currents for continuous magnetic operators},\ }\href@noop {} {\bibfield
  {journal} {\bibinfo  {journal} {Journal of Functional Analysis}\ }\textbf
  {\bibinfo {volume} {209}},\ \bibinfo {pages} {388} (\bibinfo {year}
  {2004})}\BibitemShut {NoStop}%
\bibitem [{\citenamefont {Drouot}(2021)}]{Drouot-Microlocal-2021}%
  \BibitemOpen
  \bibfield  {author} {\bibinfo {author} {\bibfnamefont {A.}~\bibnamefont
  {Drouot}},\ }\bibfield  {title} {\bibinfo {title} {Microlocal analysis of the
  bulk-edge correspondence},\ }\href@noop {} {\bibfield  {journal} {\bibinfo
  {journal} {Communications in Mathematical Physics}\ }\textbf {\bibinfo
  {volume} {383}},\ \bibinfo {pages} {2069} (\bibinfo {year}
  {2021})}\BibitemShut {NoStop}%
\bibitem [{\citenamefont {Xiao}\ \emph {et~al.}(2014)\citenamefont {Xiao},
  \citenamefont {Zhang},\ and\ \citenamefont
  {Chan}}]{Xiao-Chan-Geom-phases-1D-2014}%
  \BibitemOpen
  \bibfield  {author} {\bibinfo {author} {\bibfnamefont {M.}~\bibnamefont
  {Xiao}}, \bibinfo {author} {\bibfnamefont {Z.~Q.}\ \bibnamefont {Zhang}},\
  and\ \bibinfo {author} {\bibfnamefont {C.~T.}\ \bibnamefont {Chan}},\
  }\bibfield  {title} {\bibinfo {title} {Surface impedance and bulk band
  geometric phases in one-dimensional systems},\ }\href@noop {} {\bibfield
  {journal} {\bibinfo  {journal} {Phys. Rev. X}\ }\textbf {\bibinfo {volume}
  {4}},\ \bibinfo {pages} {021017} (\bibinfo {year} {2014})}\BibitemShut
  {NoStop}%
\bibitem [{\citenamefont {Thiang}\ and\ \citenamefont
  {Zhang}(2023)}]{Thiang-Zhang-Bulk-interface-1D-2023}%
  \BibitemOpen
  \bibfield  {author} {\bibinfo {author} {\bibfnamefont {G.~C.}\ \bibnamefont
  {Thiang}}\ and\ \bibinfo {author} {\bibfnamefont {H.}~\bibnamefont {Zhang}},\
  }\bibfield  {title} {\bibinfo {title} {Bulk-interface correspondences for
  one-dimensional topological materials with inversion symmetry},\ }\href@noop
  {} {\bibfield  {journal} {\bibinfo  {journal} {Proceedings of the Royal
  Society A}\ }\textbf {\bibinfo {volume} {479}},\ \bibinfo {pages} {20220675}
  (\bibinfo {year} {2023})}\BibitemShut {NoStop}%
\bibitem [{\citenamefont {Tsukerman}\ and\ \citenamefont
  {Markel}(2023)}]{Tsukerman-Markel-EPL2023}%
  \BibitemOpen
  \bibfield  {author} {\bibinfo {author} {\bibfnamefont {I.}~\bibnamefont
  {Tsukerman}}\ and\ \bibinfo {author} {\bibfnamefont {V.~A.}\ \bibnamefont
  {Markel}},\ }\bibfield  {title} {\bibinfo {title} {Topological features of
  {Bloch} impedance},\ }\href@noop {} {\bibfield  {journal} {\bibinfo
  {journal} {Europhysics Letters}\ }\textbf {\bibinfo {volume} {144}},\
  \bibinfo {pages} {16002} (\bibinfo {year} {2023})}\BibitemShut {NoStop}%
\bibitem [{\citenamefont {Coutant}\ and\ \citenamefont
  {Lombard}(2024)}]{Coutant-Lombard-impedance-topology-2024}%
  \BibitemOpen
  \bibfield  {author} {\bibinfo {author} {\bibfnamefont {A.}~\bibnamefont
  {Coutant}}\ and\ \bibinfo {author} {\bibfnamefont {B.}~\bibnamefont
  {Lombard}},\ }\bibfield  {title} {\bibinfo {title} {Surface impedance and
  topologically protected interface modes in one-dimensional phononic
  crystals},\ }\href@noop {} {\bibfield  {journal} {\bibinfo  {journal} {Proc.
  R. Soc. A}\ }\textbf {\bibinfo {volume} {480}} (\bibinfo {year}
  {2024})}\BibitemShut {NoStop}%
\bibitem [{\citenamefont {Felbacq}\ and\ \citenamefont
  {Rousseau}(2024)}]{Felbacq-Rousseau24}%
  \BibitemOpen
  \bibfield  {author} {\bibinfo {author} {\bibfnamefont {D.}~\bibnamefont
  {Felbacq}}\ and\ \bibinfo {author} {\bibfnamefont {E.}~\bibnamefont
  {Rousseau}},\ }\bibfield  {title} {\bibinfo {title} {Characterizing the
  topological properties of 1d non-{Hermitian} systems without the
  {Berry–Zak} phase},\ }\href@noop {} {\bibfield  {journal} {\bibinfo
  {journal} {Annalen der Physik}\ }\textbf {\bibinfo {volume} {536}},\ \bibinfo
  {pages} {2300321} (\bibinfo {year} {2024})}\BibitemShut {NoStop}%
\bibitem [{\citenamefont {Tsukerman}\ and\ \citenamefont
  {Markel}(2025)}]{Tsukerman-Markel-Topology-Bloch-impedance2024}%
  \BibitemOpen
  \bibfield  {author} {\bibinfo {author} {\bibfnamefont {I.}~\bibnamefont
  {Tsukerman}}\ and\ \bibinfo {author} {\bibfnamefont {V.~A.}\ \bibnamefont
  {Markel}},\ }\bibfield  {title} {\bibinfo {title} {Topology of {Bloch}
  impedance: {Traveling} waves, dispersive media, and electromagnetic energy},\
  }\href@noop {} {\bibfield  {journal} {\bibinfo  {journal} {J Opt}\ }\textbf
  {\bibinfo {volume} {27}},\ \bibinfo {pages} {025103} (\bibinfo {year}
  {2025})}\BibitemShut {NoStop}%
\bibitem [{\citenamefont {Yuan}\ and\ \citenamefont
  {Lu}(2006)}]{Yuan-Lu-DtN-PhC2006}%
  \BibitemOpen
  \bibfield  {author} {\bibinfo {author} {\bibfnamefont {J.}~\bibnamefont
  {Yuan}}\ and\ \bibinfo {author} {\bibfnamefont {Y.~Y.}\ \bibnamefont {Lu}},\
  }\bibfield  {title} {\bibinfo {title} {{Photonic bandgap calculations with
  Dirichlet-to-Neumann maps}},\ }\href@noop {} {\bibfield  {journal} {\bibinfo
  {journal} {J Opt Soc Am A}\ }\textbf {\bibinfo {volume} {23}},\ \bibinfo
  {pages} {3217} (\bibinfo {year} {2006})}\BibitemShut {NoStop}%
\bibitem [{\citenamefont {Hu}\ and\ \citenamefont
  {Lu}(2008)}]{Hu-Lu-DtN-PhC2008}%
  \BibitemOpen
  \bibfield  {author} {\bibinfo {author} {\bibfnamefont {Z.}~\bibnamefont
  {Hu}}\ and\ \bibinfo {author} {\bibfnamefont {Y.}~\bibnamefont {Lu}},\
  }\bibfield  {title} {\bibinfo {title} {{Improved Dirichlet-to-Neumann map
  method for modeling extended photonic crystal devices}},\ }\href@noop {}
  {\bibfield  {journal} {\bibinfo  {journal} {Opt Quant Electron}\ }\textbf
  {\bibinfo {volume} {40}},\ \bibinfo {pages} {921} (\bibinfo {year}
  {2008})}\BibitemShut {NoStop}%
\bibitem [{\citenamefont {Hu}\ and\ \citenamefont
  {Lu}(2017)}]{Hu-Lu-Bound-states2017}%
  \BibitemOpen
  \bibfield  {author} {\bibinfo {author} {\bibfnamefont {Z.}~\bibnamefont
  {Hu}}\ and\ \bibinfo {author} {\bibfnamefont {Y.~Y.}\ \bibnamefont {Lu}},\
  }\bibfield  {title} {\bibinfo {title} {Propagating bound states in the
  continuum at the surface of a photonic crystal},\ }\href@noop {} {\bibfield
  {journal} {\bibinfo  {journal} {J. Opt. Soc. Am. B}\ }\textbf {\bibinfo
  {volume} {34}},\ \bibinfo {pages} {1878} (\bibinfo {year}
  {2017})}\BibitemShut {NoStop}%
\bibitem [{\citenamefont {Zhao}\ \emph {et~al.}(2020)\citenamefont {Zhao},
  \citenamefont {Xie}, \citenamefont {Chen}, \citenamefont {Lan}, \citenamefont
  {Huang},\ and\ \citenamefont {Sha}}]{Zhao-WeiSha-Chern20}%
  \BibitemOpen
  \bibfield  {author} {\bibinfo {author} {\bibfnamefont {R.}~\bibnamefont
  {Zhao}}, \bibinfo {author} {\bibfnamefont {G.-D.}\ \bibnamefont {Xie}},
  \bibinfo {author} {\bibfnamefont {M.~L.~N.}\ \bibnamefont {Chen}}, \bibinfo
  {author} {\bibfnamefont {Z.}~\bibnamefont {Lan}}, \bibinfo {author}
  {\bibfnamefont {Z.}~\bibnamefont {Huang}},\ and\ \bibinfo {author}
  {\bibfnamefont {W.~E.~I.}\ \bibnamefont {Sha}},\ }\bibfield  {title}
  {\bibinfo {title} {First-principle calculation of {Chern} number in
  gyrotropic photonic crystals},\ }\href@noop {} {\bibfield  {journal}
  {\bibinfo  {journal} {Opt. Express}\ }\textbf {\bibinfo {volume} {28}},\
  \bibinfo {pages} {4638} (\bibinfo {year} {2020})}\BibitemShut {NoStop}%
\bibitem [{Note2()}]{Note2}%
  \BibitemOpen
  \bibinfo {note} {The computational cost for the eigenproblem \protect \eqref
  {eqn:DtN-eigenproblem-quadr} is in practice negligible, since the field at
  the interface is usually smooth and good accuracy can be attained with a
  small number of Fourier harmonics. At the same time, sparsity of FE matrices
  is easier to exploit in Dirichlet solvers than in eigensolvers with large
  matrices.}\BibitemShut {Stop}%
\bibitem [{\citenamefont {Liu}\ \emph {et~al.}(2017)\citenamefont {Liu},
  \citenamefont {Zhou}, \citenamefont {Wang},\ and\ \citenamefont
  {Gong}}]{Liu-topological-superconductors-2017}%
  \BibitemOpen
  \bibfield  {author} {\bibinfo {author} {\bibfnamefont {X.-P.}\ \bibnamefont
  {Liu}}, \bibinfo {author} {\bibfnamefont {Y.}~\bibnamefont {Zhou}}, \bibinfo
  {author} {\bibfnamefont {Y.-F.}\ \bibnamefont {Wang}},\ and\ \bibinfo
  {author} {\bibfnamefont {C.-D.}\ \bibnamefont {Gong}},\ }\bibfield  {title}
  {\bibinfo {title} {{Characterizations of topological superconductors: Chern
  numbers, edge states and Majorana zero modes}},\ }\href@noop {} {\bibfield
  {journal} {\bibinfo  {journal} {New Journal of Physics}\ }\textbf {\bibinfo
  {volume} {19}},\ \bibinfo {pages} {093018} (\bibinfo {year}
  {2017})}\BibitemShut {NoStop}%
\bibitem [{\citenamefont {Chan}(2024)}]{Chan-private2024}%
  \BibitemOpen
  \bibfield  {author} {\bibinfo {author} {\bibfnamefont {C.-T.}\ \bibnamefont
  {Chan}},\ }\href@noop {} {} (\bibinfo {year} {2024}),\ \bibinfo {note}
  {private communication}\BibitemShut {NoStop}%
\bibitem [{\citenamefont {Fukui}\ \emph {et~al.}(2005)\citenamefont {Fukui},
  \citenamefont {Hatsugai},\ and\ \citenamefont
  {Suzuki}}]{Fukui-Chern-numbers05}%
  \BibitemOpen
  \bibfield  {author} {\bibinfo {author} {\bibfnamefont {T.}~\bibnamefont
  {Fukui}}, \bibinfo {author} {\bibfnamefont {Y.}~\bibnamefont {Hatsugai}},\
  and\ \bibinfo {author} {\bibfnamefont {H.}~\bibnamefont {Suzuki}},\
  }\bibfield  {title} {\bibinfo {title} {Chern numbers in discretized
  {Brillouin} zone: {Efficient} method of computing (spin) {Hall}
  conductances},\ }\href@noop {} {\bibfield  {journal} {\bibinfo  {journal}
  {Journal of the Physical Society of Japan}\ }\textbf {\bibinfo {volume}
  {74}},\ \bibinfo {pages} {1674} (\bibinfo {year} {2005})}\BibitemShut
  {NoStop}%
\bibitem [{\citenamefont {Silva}\ \emph {et~al.}(2022)\citenamefont {Silva},
  \citenamefont {Fernandes}, \citenamefont {Morgado},\ and\ \citenamefont
  {Silveirinha}}]{Silva-Silveirinha-Tamm-states-2022}%
  \BibitemOpen
  \bibfield  {author} {\bibinfo {author} {\bibfnamefont {S.~V.}\ \bibnamefont
  {Silva}}, \bibinfo {author} {\bibfnamefont {D.~E.}\ \bibnamefont
  {Fernandes}}, \bibinfo {author} {\bibfnamefont {T.~A.}\ \bibnamefont
  {Morgado}},\ and\ \bibinfo {author} {\bibfnamefont {M.~G.}\ \bibnamefont
  {Silveirinha}},\ }\bibfield  {title} {\bibinfo {title} {Topological pumping
  and {Tamm} states in photonic systems},\ }\href@noop {} {\bibfield  {journal}
  {\bibinfo  {journal} {Phys. Rev. B}\ }\textbf {\bibinfo {volume} {105}},\
  \bibinfo {pages} {155133} (\bibinfo {year} {2022})}\BibitemShut {NoStop}%
\bibitem [{\citenamefont {Qiu}\ \emph {et~al.}(2023)\citenamefont {Qiu},
  \citenamefont {Lin}, \citenamefont {Xie},\ and\ \citenamefont
  {Zhang}}]{Qiu-Zhang-waveguide2023}%
  \BibitemOpen
  \bibfield  {author} {\bibinfo {author} {\bibfnamefont {J.}~\bibnamefont
  {Qiu}}, \bibinfo {author} {\bibfnamefont {J.}~\bibnamefont {Lin}}, \bibinfo
  {author} {\bibfnamefont {P.}~\bibnamefont {Xie}},\ and\ \bibinfo {author}
  {\bibfnamefont {H.}~\bibnamefont {Zhang}},\ }\bibfield  {title} {\bibinfo
  {title} {Mathematical theory for the interface mode in a waveguide bifurcated
  from a {Dirac} point},\ }\href@noop {} {\bibfield  {journal} {\bibinfo
  {journal} {arXiv:2304.10843}\ } (\bibinfo {year} {2023})}\BibitemShut
  {NoStop}%
\bibitem [{\citenamefont {Brillouin}(1960)}]{Brillouin60}%
  \BibitemOpen
  \bibfield  {author} {\bibinfo {author} {\bibfnamefont {L.}~\bibnamefont
  {Brillouin}},\ }\href@noop {} {\emph {\bibinfo {title} {Wave Propagation and
  Group Velocity}}}\ (\bibinfo  {publisher} {Academic Press},\ \bibinfo {year}
  {1960})\BibitemShut {NoStop}%
\bibitem [{\citenamefont {Landau}\ and\ \citenamefont
  {Lifshitz}(1984)}]{Landau84}%
  \BibitemOpen
  \bibfield  {author} {\bibinfo {author} {\bibfnamefont {L.}~\bibnamefont
  {Landau}}\ and\ \bibinfo {author} {\bibfnamefont {E.}~\bibnamefont
  {Lifshitz}},\ }\href@noop {} {\emph {\bibinfo {title} {Electrodynamics of
  Continuous Media}}}\ (\bibinfo  {publisher} {Oxford; New York: Pergamon},\
  \bibinfo {year} {1984})\BibitemShut {NoStop}%
\end{thebibliography}%

\end{document}